# WENO interpolation-based and upwind-biased schemes with free-stream preservation

Qin Li[a, b], Dong Sun[b]

[a]State Key Laboratory of Aerodynamics, Mianyang, Sichuan, 621000, China

and

[b]School of Aerospace Engineering, Xiamen University, Xiamen, Fujian, 361102, China,

Abstract Based on the understandings regarding linear upwind schemes with flux splitting to achieve free-stream preservation (Q. Li, et al. Commun. Comput. Phys., 22 (2017) 64-94), a series of WENO interpolation-based and upwind-biased nonlinear schemes are proposed in this study. By means of engagement of fluxes on midpoints, the nonlinearity of schemes is introduced through WENO interpolations, and upwind-biased features are acquired through the choice of dependent grid stencil. Regarding the third- and fifth-order versions, schemes with one and two midpoints are devised and carefully tested. With the integration of the piecewise-polynomial mapping function methods (Q. Li, et al. Commun. Comput. Phys. 18 (2015) 1417-1444), the proposed schemes are found to achieve the designed orders and free-stream preservation property. In 1-D Sod and Shu-Osher problems, all schemes succeed in yielding well predictions. In 2-D cases, the vortex preservation, supersonic inviscid flow around cylinder at $M_\infty=4$, Riemann problem and Shock-vortex interaction problems are tested. In each problem, two types of grids are employed, i.e. the uniformed/smooth grids and the randomized/partially-randomized grids. On the latter, the shock wave and complex flow structures are located/partially located. All schemes fulfill computations in uniformed/smooth grids with satisfactory results. On randomized grids, all schemes accomplish computations and yield reasonable results except the third-order one with two midpoints engaged fails in Riemann problem and shock-vortex interaction problem. Overall speaking, the proposed schemes manifest the capability to solve problems on grids with bad quality, and therefore indicate their potential in engineering applications.

Keywords: high order scheme; upwind-biased scheme; WENO; interpolation; free-stream preservation

## 1 Introduction

On developing high-order difference schemes for computations in stationary grids, large advances have been achieved since the presence of WENO schemes [1, 2]. Besides the frequently-used third- and fifth-order ones, more higher-order WENO schemes can be found in [3], and the post-processing technique to preserve positivity of density and pressure was proposed in [4]. To deal with the order degradation at critical points and improve numerical performance accordingly, the mapping function method [5-7], WENO-Z type weights with new global smoothness indicators [9-10], linear and nonlinear optimizations of weighting [11-13] were developed. Besides the WENO methodology, high-order compact schemes [14] were proposed with well-designed dispersion relations.

Although the abundant researches and outcomes, it is interesting to note that there are little applications of high-order difference schemes for engineering complexities such as airplanes, where multi-blocked grids are inevitably used, the change of grid topology usually happens, and meshes with bad quality locally might be encountered. In such situations, deformed grids would exist either within blocks or on boundaries between them. Regarding the former, Visbal and Gaitonde [15] had shown large errors would be generated in computations, and similar results were repeated by subsequent studies [16, 24]. Regarding the latter, it was reported [18] that spurious oscillations would arise around the conjunctions of grids. In cases of hypersonic computations, it is possible that strong perturbations might be invoked at the boundaries between blocks if high-order schemes like WENO are used without special treatment. Moreover, if the wall is involved in multi-block conjunctions, large errors might be generated in predictions such as heat flux and the convergence of computations would be affected. Among the causes, it is known that one of them would be the mismatches between the evaluations of grid metrics and the solving of conservative governing equations in curvilinear coordinate system.

In the derivation of conservative governing equations, theoretically zero-valued identities, i.e. metric identities (*MI*), are generated but ignored in partial differential operations [19], however they might not be cancelled out in practical computations [20]. In order to minimize the errors originated from metric evaluations, the following understandings are obtained: (1) The grid metrics are suggested to choose the conservative form other than the expanded form [21-22]; (2) The consistent form of difference scheme in each coordinate direction should be used to discretize the partial derivatives in grid metrics and governing equations [15, 23]. The key issue in above is the commutativity of difference operators [23], and the validity of which has been proved by [23] first and by [24] later. With the commutativity, it can be derived that if a uniformed flow is imposed as the initial condition, the flow field can keep constant by computations, or the so-called free-stream preservation (*FSP*) can be acquired. Although possible candidates of aforementioned consistent schemes can be linear (central or upwind), nonlinear mechanisms should be introduced to enforce numerical stability in practical applications. For linear central schemes, two methodologies have been employed in this regard, i.e. the filters [15] and WENO interpolations [17-18]. For linear upwind schemes, the choice of one fixed form might invoke numerical instability because flow characteristics usually propagate in both upwind and downwind directions. In the meanwhile, it is common in applications that the flux is split into two parts and their derivatives are discretized by corresponding windward schemes respectively. Hence, the usual treatment will make the forms of schemes inconsistent and therefore *FSP* unfulfilled.

In [25], Nonomura et al. worked on the fifth- and sixth-order WENO schemes and rearranged the original $h_{j+1/2}$ in $\frac{\partial \hat{E}}{\partial \xi} = \frac{h_{j+1/2} - h_{j-1/2}}{\Delta \xi}$ into a central consistent part plus two nonlinear dissipation parts in split fluxes. In dissipation parts a flux splitting-like treatment were proposed to replace the standard splitting, where a frozen treatment of grid metric was used. Combining the conservative metric derivation by Thomas et al. [21-22], the modified WENO schemes were shown to achieve *FSP* [25]. In [26], a detailed investigation was presented on alleviation of errors in metric evaluation regarding arbitrary linear upwind schemes with flux splitting. In the reference, the central scheme decomposition (CSD) was presented to derive the central scheme of any upwind scheme for metric evaluations, the requirement of flux splitting to offset metric-evoked errors was derived, and a demand of directionally consistent interpolation was proposed when function values on cell midpoints were involved in difference schemes. Although being used to fulfill *FSP* in above studies, the technique to interpret an upwind scheme as a central scheme plus a dissipation part is not a new stuff. As early as in periods of TVD schemes [27] and subsequent investigations on high order schemes [28-31], the technique was widely referred. In [26], some upwind linear schemes are proposed as examples, and they can only be applied in applications without shock waves. It is natural to develop their nonlinear extensions in purpose of applications.

Based on [26] and considering the introduction of nonlinearity by WENO interpolation [32] for difference schemes involving cell midpoints [17-18], a series of third-, fifth- and seventh-order WENO interpolation-based and upwind-biased schemes (abbreviated as WENOIU) are proposed in this study with the achievement of shock capturing and *FSP*. The sense of "upwind-biased" is that with respect to the symmetric point (e.g., $x_j$ for schemes of $\left(\frac{\partial}{\partial x}\right)_j$ or $x_{j+1/2}$ for aforementioned $h_{j+1/2}$), the number of upwind points (nodes and/or midpoints) is one larger than that of downwind points. From [33], such form is supposed to yield schemes with favorably less dissipation. The arrangement of the paper is arranged as follows: relevant studies are first reviewed in Section 2; then the construction of the third-, fifth-order WENOIU schemes are described in Section 3; in Section 4, the characteristics of the developed schemes are shown; in Section 5, the convergence property and *FSP* are checked; in Section 6, validating examples are tested; at last, conclusions and discussions are made in Section 7.

**2 Reviews regarding free-stream preservation**

2.1 Governing equations and grid transformation

3-D Euler equation is employed for illustration regarding *FSP* and also for validating computations. The equation in Cartesian coordinate system is

$$\frac{\partial Q}{\partial t} + \frac{\partial E}{\partial x} + \frac{\partial F}{\partial y} + \frac{\partial G}{\partial z} = 0, \tag{2.1}$$

where $Q=(\rho, \rho u, \rho v, \rho w, e)^T$, $e=p/(\gamma-1)+1/2\times\rho(u^2+v^2+w^2)$, and $\{E, F, G\}$ are inviscid fluxes. It is usual that not Eq. (2.1) but its transformation in curvilinear coordinate system ($\xi, \eta, \zeta$) is solved in computations unless uniformed grids are used. The conversion from $(x, y, z)$ to $(\xi, \eta, \gamma)$ is derived through the chain rule such as: $\partial/\partial x = \xi_x\left(\partial/\partial \xi\right) + \eta_x\left(\partial/\partial \eta\right) + \zeta_x\left(\partial/\partial \zeta\right)$, and the transformed equation is

$$\frac{\partial \hat{Q}}{\partial t} + \frac{\partial \hat{E}}{\partial \xi} + \frac{\partial \hat{F}}{\partial \eta} + \frac{\partial \hat{G}}{\partial \zeta} = 0, \tag{2.2}$$

where $\hat{Q} = J^{-1}Q$ with $J^{-1} = \left|\frac{\partial(x,y,z)}{\partial(\xi,\eta,\zeta)}\right|$, $\hat{E} = (\hat{\xi}_x E + \hat{\xi}_y F + \hat{\xi}_z G)$ with $\hat{\xi}_x = J^{-1}\xi_x$ and etc. In the derivation, Leibniz formula is freely used and the theoretically zero-valued term ($\hat{E}\cdot I_x + \hat{F}\cdot I_y + \hat{G}\cdot I_z$) has been dropped, where the metric identity $I_{x_i}$ is defined as $I_{x_i} = \left(\hat{\xi}_{x_i}\right)_\xi + \left(\hat{\eta}_{x_i}\right)_\eta + \left(\hat{\zeta}_{x_i}\right)_\zeta$. Although partial differential operators such as $\partial/\partial \xi$ in Eq. (2.2) are mathematically indistinguishable among themselves, they might be numerically different depending on implementations. One concern regarding the above issue is usually invoked, i.e. when the uniform flow condition is imposed in free space, whether $Q$ would keep constant or *FSP* would be fulfilled or not. To achieve *FSP*, it is known [23-24] that conservative forms of grid metrics and the consistent use of difference schemes in each direction should be chosen when discretizing derivatives in metrics and fluxes in Eq. (2.2).

2.2 Conservative formulations of grid metrics

Taking $\hat{\xi}_x$ for example, the conservative form [21] is:

$$\hat{\xi}_x = (y_\eta z)_\zeta - (y_\zeta z)_\eta, \quad (2.3)$$

while the symmetric conservative form [22] is

$$\hat{\xi}_x = \tfrac{1}{2}\left[(y_\eta z)_\zeta - (y_\zeta z)_\eta + (z_\zeta y)_\eta - (z_\eta y)_\zeta \right]. \quad (2.4)$$

The complete formulations are suggested to [26].

Although not theoretically relating with metric-caused errors, the evaluations of Jacobian are suggested as [35]

$$J^{-1} = \tfrac{1}{3}\left[\hat{\xi}_x x_\xi + \hat{\xi}_y y_\xi + \hat{\xi}_z z_\xi + \hat{\eta}_x x_\eta + \hat{\eta}_y y_\eta + \hat{\eta}_z z_\eta + \hat{\zeta}_x x_\zeta + \hat{\zeta}_y y_\zeta + \hat{\zeta}_z z_\zeta \right] \quad (2.5)$$

or

$$J^{-1} = \tfrac{1}{3}\left[(\hat{\xi}_x x)_\xi + (\hat{\xi}_y y)_\xi + (\hat{\xi}_z z)_\xi + (\hat{\eta}_x x)_\eta + (\hat{\eta}_y y)_\eta + (\hat{\eta}_z z)_\eta + (\hat{\zeta}_x x)_\zeta + (\hat{\zeta}_y y)_\zeta + (\hat{\zeta}_z z)_\zeta \right], \quad (2.6)$$

which is shown by [26] to be the straightforward outcome of the formula

$$\nabla \cdot \vec{A} = \tfrac{1}{J^{-1}}\left[ \left(\hat{\xi}_x,\hat{\xi}_y,\hat{\xi}_z\right)^T \cdot \vec{A}_\xi + \left(\hat{\eta}_x,\hat{\eta}_y,\hat{\eta}_z\right)^T \cdot \vec{A}_\eta + \left(\hat{\zeta}_x,\hat{\zeta}_y,\hat{\zeta}_z\right)^T \cdot \vec{A}_\zeta \right] \quad \text{or}$$

$$\nabla \cdot \vec{A} = \tfrac{1}{J^{-1}}\left\{ \left[\left(\hat{\xi}_x,\hat{\xi}_y,\hat{\xi}_z\right)^T \cdot \vec{A}\right]_\xi + \left[\left(\hat{\eta}_x,\hat{\eta}_y,\hat{\eta}_z\right)^T \cdot \vec{A}\right]_\eta + \left[\left(\hat{\zeta}_x,\hat{\zeta}_y,\hat{\zeta}_z\right)^T \cdot \vec{A}\right]_\zeta \right\} \quad \text{by letting}$$

$\vec{A} = \vec{r}$. Besides, the derivations of grid metrics in $J^{-1}$ are suggest to take Eq. (2.3) or (2.4).

2.3 Difference schemes to achieve free-stream preservation

In order to achieve free-stream preservation, it is proven [23-24] that difference schemes should be the same in discretizing derivatives in grid metrics and Eq. (2.2) in each grid direction while combining with the use of Eq. (2.3) or (2.4) [15, 23-24]. It is obvious that central schemes can fulfill this requirement no matter flux splitting is applied or not. Moreover, when midpoints are adopted in schemes, WENO interpolation can be employed to derive function values on them with the engagement of flux splitting naturally, and through which necessary dissipations will be invoked when shock waves are encountered [18].

For upwind schemes, a method in [25] was proposed by rearranging the canonical fifth-order WENO schemes into one consistent central part plus two nonlinear dissipation parts, where the central part is used for metric evaluation and metric linearization is used for flux splitting in dissipation parts. In [26], investigations are carried out independently as well. It reads that [26], if a $r$th-order upwind scheme $\delta^+$ for $\partial f$ at $x_j$ is considered (here "$\partial$" denotes the differential operator for simplicity), its counterpart $\delta^-$ in downwind direction will be defined accordingly, therefore a central difference scheme $\delta^{c,(1)}$ can be derived as

$$\left(\delta^{c,(1)} f\right) = \left[\left(\delta^+ f_j\right) + \left(\delta^- f_j\right)\right]/2. \tag{2.7}$$

From the equation, a central discretization of $\partial^{(2\lfloor r/2 \rfloor + 2)} f$ naturally comes out through $\delta^{c,(2\lfloor r/2 \rfloor + 2)} = \left(\delta^{c,(1)} f\right)_j - \left(\delta^+ f\right)_j$ where the number in bracket superscripted such as "$2\lfloor r/2 \rfloor + 2$" denotes the order. Through the above procedure, the following decompositions are straightforward

$$\begin{cases} \left(\delta^+ f_j\right) = \left(\delta^{c,(1)} f\right) - \left(\delta^{c,(2\lfloor r/2 \rfloor + 2)} f\right) \\ \left(\delta^- f_j\right) = \left(\delta^{c,(1)} f\right) + \left(\delta^{c,(2\lfloor r/2 \rfloor + 2)} f\right) \end{cases} \tag{2.8}$$

which is referred as central scheme decomposition in [26].

As mentioned in the introduction, the above practice has already been done by previous studies [27-31]. For example, it has been pointed out in [31] that a fifth-order upwind scheme could be derived by adding the same order dissipation/filter to a sixth-order-maximum scheme as

$$\begin{aligned} &\beta f'_{i-2} + \alpha f'_{i-1} + f'_i + \alpha f'_{i+1} + \beta f'_{i+2} \\ &= \tfrac{1}{\Delta x}\left[a_3\left(f_{i+3} - f_{i-3}\right) + a_2\left(f_{i+2} - f_{i-2}\right) + a_1\left(f_{i+1} - f_{i-1}\right)\right] + \\ &\quad \tfrac{1}{\Delta x}\left[b_3\left(f_{i+3} + f_{i-3}\right) + b_2\left(f_{i+2} + f_{i-2}\right) + b_1\left(f_{i+1} + f_{i-1}\right) + b_0 f_i\right] \end{aligned} \tag{2.9}$$

where $\alpha$, $\beta$, $a_i$, $b_i$ are coefficients to satisfy order requirement. It is easy to see that when $\alpha=\beta=0$, $a_3=-b_3=1/60$, $a_2=-9/60$, $b_2=6/60$, $a_1=45/60$, $b_1=-15/60$ and $b_0=20/60$, the linear WENO5 will be acquired as

$$\begin{aligned} f'_i &= \tfrac{1}{60\Delta x}\left(-f_{i-3} + 9 f_{i-2} - 45 f_{i-1} + 45 f_{i+1} - 9 f_{i+2} + f_{i+3}\right) + \\ &\quad \tfrac{1}{60\Delta x}\left(-f_{i-3} + 6 f_{i-2} - 15 f_{i-1} + 20 f_i - 15 f_{i+1} + 6 f_{i+2} - f_{i+3}\right). \\ &= \tfrac{1}{60\Delta x}\left(-2 f_{i-3} + 15 f_{i-2} - 60 f_{i-1} + 20 f_i + 30 f_{i+1} - 3 f_{i+2}\right) \end{aligned} \tag{2.10}$$

In [26], the derived $\delta^{c,(1)}$ is used to discretize grid metrics, while the original $\delta^+$ and $\delta^-$ will be normally used in discretizing flux derivatives as usual. Besides above manipulations, another requirement regarding the flux splitting necessary to attain *FSP* is proposed in [26]. Considering that the flux splitting with the form as $\hat{E}^\pm = \tfrac{1}{2}\left(\hat{E} \pm \hat{A} \cdot Q\right)$ or $\hat{E}^\pm = \tfrac{1}{2}\left(\hat{E} \pm \hat{E}_{ref}\right)$, one sufficient condition is: when the uniformed-flow condition is imposed, $\hat{A}$ and $\hat{E}_{ref}$ should be locally constant at least at the dependent grid stencil of $\delta^{c,(2\lfloor r/2 \rfloor + 2)}$. Details of the analysis can be found in [26], and the form of $\hat{A}$ used in this paper is: $\hat{A} = \alpha \cdot MAX(\hat{U} + c|\hat{k}|)$ where $\hat{U} = \hat{k}_x u + \hat{k}_y v + \hat{k}_z w$, $\hat{k} = \left(\hat{k}_x^2 + \hat{k}_y^2 + \hat{k}_z^2\right)^{1/2}$ with $k$ denoting one curvilinear coordinate and $\alpha=1.1\sim1.2$, and where *MAX* runs over the full range of the local $k$ direction.

At last, it is worthwhile to mention that the conservative form of a linear scheme is convertible to its expanded form. For example, the conservative linear WENO5 by $h_{j+1/2}$ in $\delta f_j = (h_{j+1/2} - h_{j-1/2})/\Delta x$ is

$$\begin{aligned} h_{j+1/2} &= \tfrac{1}{60\Delta x}\left(f_{j-2} - 8f_{j-1} + 37f_j + 37f_{j+1} - 8f_{j+2} + f_{j+3}\right) \\ &+ \tfrac{1}{60\Delta x}\left(f_{j-2} - 5f_{j-1} + 10f_j - 10f_{j+1} + 5f_{j+2} - f_{j+3}\right), \\ &= \tfrac{1}{60\Delta x}\left(2f_{j-2} - 13f_{j-1} + 47f_j + 27f_{j+1} - 3f_{j+2}\right) \end{aligned} \tag{2.11}$$

which is just equivalent to Eq. (2.10).

2.4 Interpolations to achieve free-stream preservation when cell midpoints are engaged

When cell midpoints are engaged in difference schemes discussed in previous section, two types of interpolations will be invoked. The first one regards the derivation of function values on midpoints. In this aspect, WENO interpolations can be applied, and through which the well-known WCNS series scheme are established [17]. In the interpolations, the following types of variable are observed for usage, i.e. the primitive, conservative, characteristic variables [36] and the flux [37]. Using WCNS, the first three types have been tested to achieve *FSP* [17-18] on deformed grids such as randomized ones; for the last one, although successful computations have been made on typical 2-D problems on uniformed Cartesian grids, it would be uncertain to achieve *FSP* on deformed grids because grid metrics are incorporated in flux during interpolation. Although metrics on midpoints can also be derived by interpolation, such derivation will be related to interpolation in specific direction, and the validity of which might be subjected to doubts. Hence, in order to achieve *FSP*, the first three types of variables are suggested to use.

The second type of interpolation regards grid metrics at cell midpoints. Required by difference schemes [17-18] engaging midpoints, grid metrics therein should be acquired for computation. The most convenient way for acquisition is that: evaluate the function values at nodes firstly by following the instructions in Section 2.2 and 2.3, then interpolate them to the midpoints. In [26], a sufficient requirement to achieve *FSP* has been proposed as: the consistent linear interpolation should be casted on each coordinate direction (so-called directionally consistent interpolation) in the evaluation of metrics and those at midpoints in fluxes, while the interpolations could be different in different directions. In the reference, a proof of the requirement has been given and numerical validations have been provided.

**3 WENO interpolation-based and upwind-biased scheme with free-stream preservation**

Based on understandings in Section 2, a series of third- and fifth-order upwind-biased conservative schemes (WENOIU) are proposed with the nonlinearity introduced by WENO interpolation, where the number of grid points (nodes and midpoints) in upwind direction is equal or one larger than that in downwind direction with respect to $x_{j+1/2}$. Besides of the achievement of *FSP*, proposed schemes are expected to solve problems with shock waves and therefore indicate engineering potentials.

3.1 Construction of difference schemes

Consider one-dimensional hyperbolic conservative law for illustration,

$$u_t + f(u)_x = 0. \tag{3.1}$$

Supposing the grid is discretized as $x_i = i \cdot \Delta x$ where $\Delta x$ is the uniformed space interval, the semi-discretized conservative formulation at $x_j$ holds

$$(u_t)_j = -\left(\hat{f}_{j+1/2} - \hat{f}_{j-1/2}\right)/\Delta x, \tag{3.2}$$

where $\hat{f}(u)$ is implicitly defined by $f(x) = \frac{1}{\Delta x}\int_{x-\Delta x/2}^{x+\Delta x/2} \hat{f}(x')dx'$. The formulation in [34] to derive $\hat{f}(u)$ is chosen in this study to develop WENOIU schemes as

$$\hat{f}_{j+1/2} = f_{j+1/2} - \tfrac{1}{24}\Delta x^2 \left(\tfrac{\partial^2 f}{\partial x^2}\right)_{j+1/2} + \tfrac{7}{5760}\Delta x^4 \left(\tfrac{\partial^4 f}{\partial x^4}\right)_{j+1/2} - \tfrac{31}{967680}\Delta x^6 \left(\tfrac{\partial^6 f}{\partial x^6}\right)_{j+1/2} + O(\Delta x^8). \tag{3.3}$$

Considering similar difference schemes involving cell midpoints such as that in [14], it is conceivable that high-order upwind schemes can be derived through discretizing $\left(\tfrac{\partial^n f}{\partial x^n}\right)_{j+1/2}$.

It is trivial that two types of function values can be used to discretize derivatives in Eq. (3.3), i.e. the values on grid nodes and/or that on cell midpoints, and in order to introduce nonlinearity, at least one midpoint should be used in Eq. (3.3). In this study, practices are made to construct schemes with various midpoints engaged.

3.2 Third-order WENOIU schemes

Considering Eq. (3.3), it is obvious that only $\left(\tfrac{\partial^2 f}{\partial x^2}\right)_{j+1/2}$ needs to be discretized.

(1) Scheme with only one midpoint $x_{j+1/2}$ engaged

It is easy to obtain the approximation of $\hat{f}_{j+1/2}$, or $h_{j+1/2}$ as

$$h_{j+1/2} = f_{j+1/2} + \tfrac{1}{24}\left(-f_{j-1} + 2f_j - f_{j+1}\right). \tag{3.4}$$

Considering the upwind-biased nature, the number of nodes to approximate the derivative should be $2n+1$ with $n \geq 1$, and therefore the total point number including that of $f_{j+1/2}$ will be $2n+2$. Hence, the above scheme should have one free parameter. Based on the understanding, the more general form can be derived as

$$h_{j+1/2} = \alpha f_{j+1/2} + \left(\tfrac{1}{8}\alpha - \tfrac{1}{6}\right)f_{j-1} + \left(-\tfrac{3}{4}\alpha + \tfrac{5}{6}\right)f_j + \left(-\tfrac{3}{8}\alpha + \tfrac{1}{3}\right)f_{j+1}. \tag{3.5}$$

and when $\alpha = 8/9$, the non upwind-biased scheme with minimum points can be obtained as

$$h_{j+1/2} = \tfrac{8}{9}f_{j+1/2} - \tfrac{1}{18}f_{j-1} + \tfrac{1}{6}f_j. \tag{3.6}$$

The accuracy relation of Eq. (3.5) is:

$$f_{j+1/2} - \tfrac{1}{24}\Delta x^2\left(\tfrac{\partial^2 f}{\partial x^2}\right)_{j+1/2} + \left(\tfrac{1}{16}\alpha - \tfrac{1}{12}\right)\Delta x^3\left(\tfrac{\partial^3 f}{\partial x^3}\right)_{j+1/2} + \left(\tfrac{3}{128}\alpha - \tfrac{37}{1152}\right)\Delta x^4\left(\tfrac{\partial^4 f}{\partial x^4}\right)_{j+1/2} + O(\Delta x^5).$$

Taking the order and number of midpoints as indicators, the scheme by Eq. (3.5) is referred as WENOIU3-1MP where "MP" denotes midpoints. In this study, only Eq. (3.4) is tested while the use of Eq. (3.5) will be further discussed in other investigations.

According to Eq. (2.7), $\delta^{c,(1)}f$ of Eq. (3.4) to compute grid metrics is

$$\delta^{c,(1)}f = \frac{1}{\Delta x}\left[\left(f_{j+1/2} - f_{j-1/2}\right) + \tfrac{1}{48}\left(-f_{j+2} + 2f_{j+1} - 2f_{j-1} + f_{j-2}\right)\right], \tag{3.7}$$

which is a fourth-order central difference scheme.

(2) Scheme with two midpoints engaged

The scheme can be derived as

$$h_{j+1/2} = f_{j+1/2} + \tfrac{1}{6}\left(-f_{j-1/2} + 2f_j - f_{j+1/2}\right) \tag{3.8}$$

with the accuracy relation as

$$f_{j+1/2} - \tfrac{1}{24}\Delta x^2 \left(\tfrac{\partial^2 f}{\partial x^2}\right)_{j+1/2} + \tfrac{1}{48}\Delta x^3 \left(\tfrac{\partial^3 f}{\partial x^3}\right)_{j+1/2} - \tfrac{7}{1152}\Delta x^4 \left(\tfrac{\partial^4 f}{\partial x^4}\right)_{j+1/2} + O(\Delta x^5).$$

Similarly, the scheme by Eq. (3.8) is referred as WENOIU3-2MP.

According to Eq. (2.7), the fourth-order $\delta^{c,(1)}f$ of Eq. (3.8) to compute grid metrics is

$$\delta^{c,(1)}f = \frac{1}{12\Delta x}\left[2\left(f_{j+1} - f_{j-1}\right) + \left(-f_{j+3/2} + 11f_{j+1/2} - 11f_{j-1/2} + f_{j-3/2}\right)\right] \tag{3.9}$$

(3) WENO interpolation [32]

In Eqns. (3.4)-(3.6) and (3.8), unknown fluxes on half nodes should be acquired by interpolation, and through which the implementation of scheme could be accomplished. The third-order WENO interpolation from [32] is realized as

$$u_{j+1/2} = \sum_{k=0}^{r-1} \omega_k p_k^r \tag{3.10}$$

where $r$ denotes the grid number of stencil of candidate scheme ($r=2$ here) and

$$p_k^r = \sum_{l=0}^{r-1} b_{k,l}^r u_{j-r+k+l+1}. \tag{3.11}$$

The coefficients $b_{k,l}^r$ in Eq. (3.11) are given in Table 1 [32]. $\omega_k$ is the nonlinear weight derived from its linear counterpart $C_k^r$ which is shown in Table 1 as well. The canonical weighting procedure is

$$\omega_k = \alpha_k \bigg/ \sum_{l=0}^{r-1} \alpha_l \tag{3.12}$$

where

$$\alpha_k = C_k^r \big/ \left(\varepsilon + IS_k\right)^2 \tag{3.13}$$

and

$$IS_k = \left(u_{j+k} - u_{j+k-1}\right)^2. \tag{3.14}$$

It is known that at critical points, order degradation of standard WENO interpolation will occur. If such occurrence is quite concerned, methods such as the mapping function [5] can be applied, and a specific one named as the piecewise polynomial mapping function (abbreviated as PPM) [7] is chosen in this study. In case of third-order scheme, the second-order PPM is needed and has the form as

$$g(\omega) = \begin{cases} C_k \left[1 - \left(\frac{\omega}{C_k} - 1\right)^2\right] & \omega \le C_k \\ C_k - \frac{1}{C_k - 1}(\omega - C_k)^2 & \omega > C_k \end{cases}. \quad (3.15)$$

The procedure of usage is that after initial evaluation of $\omega$ by Eqns. (3.12) - (3.14), Eq. (3.15) is invoked for $\omega$, then the normalization is executed on mapped values to yield the final $\omega$. From [5, 7], it can be conceived that when Eq. (3.15) is used, $\varepsilon$ can have a very small quantity such as $10^{-40}$, otherwise it takes the value $10^{-6}$ in the computation.

**Table 1. Coefficients $b_{k,l}^r$ and $C_k^r$ in interpolation schemes**

| r | k | $C_k^r$ | $b_{k,l}^r$ | | |
|---|---|---|---|---|---|
| | | | l=0 | l=1 | l=2 |
| 2 | 0 | 1/4 | -1/2 | 3/2 | - |
| | 1 | 3/4 | 1/2 | 1/2 | - |
| 3 | 0 | 1/16 | 3/8 | -10/8 | 15/8 |
| | 1 | 10/16 | -1/8 | 6/8 | 3/8 |
| | 2 | 5/16 | 3/8 | 6/8 | -1/8 |

3.3 Fifth-order WENOIU schemes

In this situation, both $\left(\frac{\partial^2 f}{\partial x^2}\right)_{j+1/2}$ and $\left(\frac{\partial^4 f}{\partial x^4}\right)_{j+1/2}$ in (3.3) need to be discretized.

(1) Scheme with only one midpoint $f_{j+1/2}$ engaged

By discretizing the second- and fourth-order derivatives, the following fifth-order scheme can be obtained as

$$h_{j+1/2} = f_{j+1/2} + \tfrac{19}{1920} f_{j-2} - \tfrac{29}{480} f_{j-1} + \tfrac{77}{960} f_j - \tfrac{3}{160} f_{j+1} - \tfrac{7}{640} f_{j+2}, \quad (3.16)$$

As mentioned before, the more general form which has one free parameter is

$$h_{j+1/2} = \alpha f_{j+1/2} + \left(-\tfrac{3}{128}\alpha + \tfrac{1}{30}\right) f_{j-2} + \left(\tfrac{5}{32}\alpha - \tfrac{13}{60}\right) f_{j-1} + \left(-\tfrac{45}{64}\alpha + \tfrac{47}{60}\right) f_j \\ + \left(-\tfrac{15}{32}\alpha + \tfrac{9}{20}\right) f_{j+1} + \left(\tfrac{5}{128}\alpha - \tfrac{1}{20}\right) f_{j+2} \quad (3.17)$$

When $\alpha=32/25$, the non upwind-biased scheme with one point less is obtained as:

$$h_{j+1/2} = \tfrac{32}{25} f_{j+1/2} + \tfrac{1}{300} f_{j-2} - \tfrac{1}{60} f_{j-1} + \tfrac{7189}{24000} f_j - \tfrac{3}{20} f_{j+1} \quad (3.18)$$

The accuracy relation of Eq. (3.17) is

$$f_{j+1/2} - \tfrac{1}{24}\Delta x^2 \left(\tfrac{\partial^2 f}{\partial x^2}\right)_{j+1/2} + \tfrac{7}{5760}\Delta x^4 \left(\tfrac{\partial^4 f}{\partial x^4}\right)_{j+1/2} + \tfrac{1}{3840}(45\alpha - 64)\Delta x^5 \left(\tfrac{\partial^5 f}{\partial x^5}\right)_{j+1/2} \\ - \tfrac{1}{138240}(675\alpha - 983)\Delta x^6 \left(\tfrac{\partial^6 f}{\partial x^6}\right)_{j+1/2} + O(\Delta x^7)$$

.

The scheme by Eq. (3.17) is referred as WENOIU5-1MP. As before, only Eq. (3.16) is tested in current study.

According to Eq. (2.7), the sixth-order $\delta^{c,(1)}f$ of Eq. (3.16) is

$$\delta^{c,(1)}f = \frac{1}{\Delta x}\left(f_{j+1/2} - f_{j-1/2}\right) + \frac{1}{\Delta x}\left[\begin{array}{l}\frac{19}{3840}\left(f_{j+3} - f_{j-3}\right) - \frac{13}{320}\left(f_{j+2} - f_{j-2}\right) + \\ \frac{17}{256}\left(f_{j+1} - f_{j-1}\right)\end{array}\right]. \quad (3.19)$$

(2) Scheme with two midpoints engaged

The scheme can be derived as

$$h_{j+1/2} = f_{j+1/2} + \frac{1}{30}\left[\left(-4f_{j+1} + f_j + f_{j-1}\right) + \left(6f_{j+1/2} - 4f_{j-1/2}\right)\right] \quad (3.20)$$

with the accuracy relation as

$$f_{j+1/2} - \tfrac{1}{24}\Delta x^2 \left(\tfrac{\partial^2 f}{\partial x^2}\right)_{j+1/2} + \tfrac{7}{5760}\Delta x^4 \left(\tfrac{\partial^4 f}{\partial x^4}\right)_{j+1/2} - \tfrac{1}{960}\Delta x^5 \left(\tfrac{\partial^5 f}{\partial x^5}\right)_{j+1/2} + \tfrac{47}{138240}\Delta x^6 \left(\tfrac{\partial^6 f}{\partial x^6}\right)_{j+1/2} + O(\Delta x^7).$$

Similarly, the scheme by Eq. (3.20) is denoted as WENOIU5-2MP. The sixth-order of $\delta^{c,(1)}f$ of Eq. (3.20) by Eq. (2.7) is

$$\delta^{c,(1)}f = \frac{1}{60\Delta x}\left(f_{i+2} - 4f_{j+1} + 4f_{j-1} - f_{j-2}\right) + \frac{1}{15\Delta x}\left(-f_{j+3/2} + 19f_{j+1/2} - 19f_{j-1/2} + f_{j-3/2}\right). \quad (3.21)$$

(3) WENO interpolation [32]

The function values at cell midpoints are derived through WENO interpolation as that in Section 3.2 with $r=3$. The coefficients in Eqns. (3.10) - (3.11) are shown in Table 1 as well. The smoothness indicators in Eq. (3.13) are

$$\begin{cases} IS_0 = \tfrac{1}{3}\left(4u_{j-2}^2 - 19u_{j-2}u_{j-1} + 25u_{j-1}^2 + 11u_{j-2}u_j - 31u_{j-1}u_j + 10u_j^2\right) \\ IS_1 = \tfrac{1}{3}\left(4u_{j-1}^2 - 13u_{j-1}u_j + 13u_j^2 + 5u_{j-1}u_{j+1} - 13u_j u_{j+1} + 4u_{j+1}^2\right) \\ IS_2 = \tfrac{1}{3}\left(10u_j^2 - 31u_j u_{j+1} + 25u_{j+1}^2 + 11u_j u_{j+2} - 19u_{j+1}u_{j+2} + 4u_{j+2}^2\right) \end{cases} \quad (3.22)$$

The Taylor series expansion of Eq. (3.22) is: $\left(\tfrac{\partial u}{\partial x}\right)_j^2 \Delta x^2 + \left(a \tfrac{\partial u}{\partial x}\tfrac{\partial^3 u}{\partial x^3} + \tfrac{13}{12}\left(\tfrac{\partial^2 u}{\partial x^2}\right)^2\right)_j \Delta x^4 + O(\Delta x^5)$ where $a$=-2/3, 1/3 and -2/3 respectively. When $\left(\partial u/\partial x\right)_j = 0$, $O(\Delta x^5)$ is not zero usually, therefore Eq. (3.22) still suffers from order degradation at critical points. If such situation is quite concerned, the mapping function of [5] or the following third-order PPM can be used [7]

$$g(\omega) = \begin{cases} C_k\left[1 + \left(\tfrac{\omega}{C_k} - 1\right)^3\right] & \omega \leq C_k \\ C_k + \left(\tfrac{1}{C_k - 1}\right)^2 (\omega - C_k)^3 & \omega > C_k \end{cases}. \quad (3.23)$$

3.4 Variables used in WENO interpolation

As mentioned in Section 2.4, the primitive, conservative and characteristic variables can be chosen in WENO interpolations. From literatures [17, 36], the last type was supposed to yield results with less oscillations usually in smooth or uniformed grid. The key issue in the use of characteristic variables is to define proper flux and to derive its Jacobian accordingly, through which the left eigenvectors of the Jacobian are used to project conservative vectors from component space into characteristic field and the right eigenvectors are used to project results of interpolations back. On consideration of *FSP*, the following procedure is chosen to construct aforementioned flux at $j+1/2$ as: the variables and metrics are acquired separately at $j+1/2$ as following, afterwards they are combined to form the flux required. Considering the case of discretizing $\partial \hat{E}/\partial \xi$ in Eq. (2.2) as an example, the referred flux tagged at $j+1/2$ would be

$$\hat{\xi}'_{x,j+1/2}E(Q'_{j+1/2}) + \hat{\xi}'_{y,j+1/2}F(Q'_{j+1/2}) + \hat{\xi}'_{z,j+1/2}G(Q'_{j+1/2})$$ where $Q'_{j+1/2} = Q_j$ for $h^+_{j+1/2}$ and $Q'_{j+1/2} = Q_{j+1}$ for $h^-_{j+1/2}$, and where $\hat{\xi}'_{x,j+1/2} = \left(\hat{\xi}_{x,j} + \hat{\xi}_{x,j+1}\right)/2$ for simplicity. Practices show that such implementation will yield results with less oscillations when solving problems with shock waves, in the meanwhile *FSP* will not be affected. With regard to the variables to be projected [17, 36], both primitive and conservative variables can be used, and the latter is chosen in this study.

Thus far, the constructions of WENOIU3 and WENOIU5 are accomplished. For completeness, the formulations of seventh-order WENOIU schemes are provided in Appendix.

**4 Spectral features of proposed schemes**

To explore the mathematical characteristics of proposed schemes, two analyses are carried out, i.e. the standard Fourier analysis and the one of Approximate Dispersion Relation (ADR).

4.1 Fourier analysis

In [38], Fourier transformation was employed to show the spectral characteristics of a difference scheme. Consider the following generalized form of scheme:

$$\left(\frac{\partial f(x)}{\partial x}\right)_j \approx \frac{1}{\Delta x}\left(\sum_{l_1=-N_1}^{M_1} a_{l_1} f_{j+l_1} + \sum_{l_2=-N_2}^{M_2} b_{l_2} f_{j+1/2+l_2}\right). \tag{4.1}$$

When $x_j$ is treated as the continuous variable $x$, Fourier transformation can be applied through $\tilde{f}(k) = \frac{1}{2\pi}\int_{-\infty}^{\infty} f(x)e^{-ikx}dx$ where $k$ is the wave number. Usually, the scaled wave number $\kappa = k\Delta x$ is used in analysis other than the original $k$, and a modified $\kappa'$ with regard to Eq. (4.1) can be derived as [38]

$$\kappa' = -i\left(\sum_{l_1=-N_1}^{M_1} a_{l_1} e^{il_1\kappa} + \sum_{l_2=-N_2}^{M_2} b_{l_2} e^{i(l_2+1/2)\kappa}\right). \tag{4.2}$$

Through Eq. (4.2), the real and imaginary parts of $\kappa'$ of above schemes can be obtained and are summarized in Table 2. Because the interpolation algorithms are relatively independent from the schemes, they are not incorporated into spectral relations in the table temporarily. For comparison, two additional central schemes (CS) based on midpoints by [14] are also considered, i.e. the fourth-order $f'_j = \frac{1}{\Delta x}\left[\frac{9}{8}\left(f_{j+1/2} - f_{j-1/2}\right) - \frac{1}{24}\left(f_{j+3/2} - f_{j-3/2}\right)\right]$ and the sixth-order $f'_j = \frac{1}{\Delta x}\left[\frac{75}{64}\left(f_{j+1/2} - f_{j-1/2}\right) - \frac{25}{384}\left(f_{j+3/2} - f_{j-3/2}\right) + \frac{3}{640}\left(f_{j+5/2} - f_{j-5/2}\right)\right]$.

Based on the formulations in Table 2, the spectral characteristics of discussed schemes in linear forms are drawn in Fig. 1. The schemes are classified as two groups according to the accuracy order, i.e. WENOIU3-1MP at $\alpha=1$, WENOIU3-2MP and CS4 as the first group in Fig. 1a, and WENOIU5-1MP at $\alpha=1$, WENOIU5-2MP and CS6 as the second group in Fig. 1.b. In each group, results of schemes only and schemes with linear interpolations incorporated (denoted by the tag "-intp") are shown. In the first group in Fig. 1a, when difference schemes only are concerned, CS4 shows the best performance in dispersion relation while WENOIU3-2M behaves the most dissipative; when linear interpolations are incorporated, the relative relationship of spectrum preserves in WENOIU3 schemes, while CS4 shows the least dissipative and the reduced performance on dispersion. In the second group in Fig. 1b, when difference schemes only are concerned, WENOIU5-2MP and CS4 show better dispersion performance than that of WENOIU5-1MP; when linear interpolations are incorporated, all schemes show similar dispersion performance, and it is interesting to note CS6 indicates a dissipation larger than that of WENOIU5-1MP and WENOIU5-2MP behaves the most dissipative. Such interpolation-caused phenomenon might need further analysis. Overall, the proposed schemes indicate certain amount of dissipation which is preferable for numerical stability.

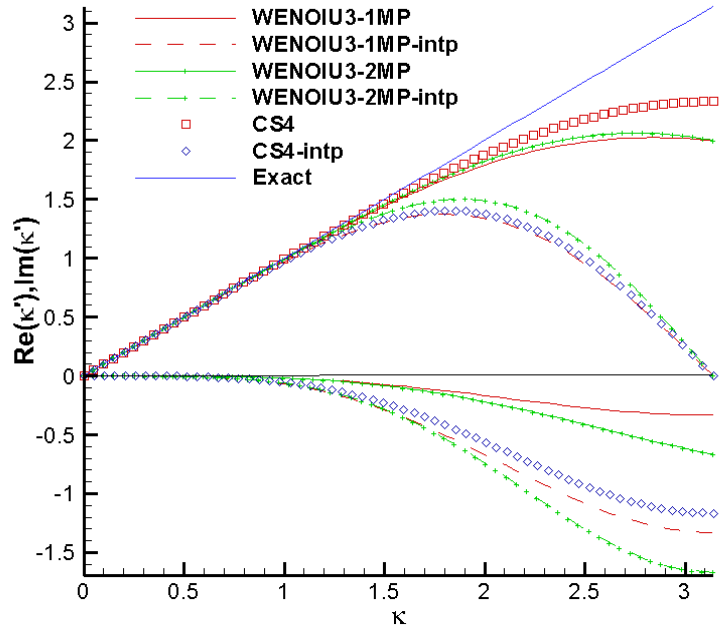

(a) WENOIU3-1MP, -2MP, CS4 and CS6

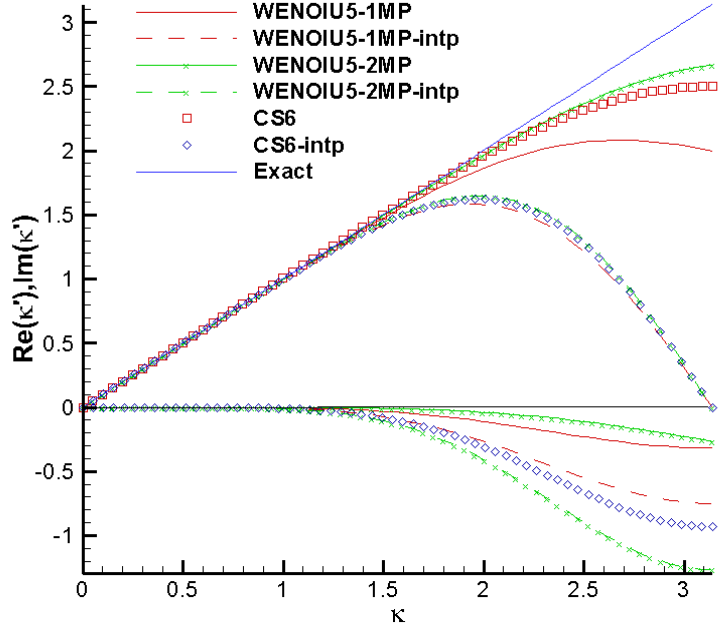

(b) WENOIU5-1MP, -2MP, CS4 and CS6

Fig.1 Dispersion and dissipation relations of difference schemes by Fourier transformation

**Table 2 Dispersion and dissipation relations of difference schemes**

| Scheme | | Spectral relation |
|---|---|---|
| WENOIU3-1MP | Dispersion | $2\alpha \sin(\frac{1}{2}\kappa) + (\frac{4}{3} - \frac{5}{4}\alpha)\sin(\kappa) + (-\frac{1}{6} + \frac{1}{8}\alpha)\sin(2\kappa)$ |
| | Dissipation | $(-\frac{1}{3} + \frac{1}{4}\alpha)\cos^2(\kappa) + (\frac{2}{3} - \frac{1}{2}\alpha)\cos(\kappa) + (-\frac{1}{3} + \frac{1}{4}\alpha)$ |
| WENOIU3-2MP | Dispersion | $\frac{11}{6}\sin(\frac{1}{2}\kappa) - \frac{1}{6}\sin(\frac{3}{2}\kappa) + \frac{1}{3}\sin(\kappa)$ |
| | Dissipation | $\frac{2}{3}\sin(\frac{1}{2}\kappa)^2(\cos(\frac{1}{2}\kappa) - 1)$ |
| WENOIU5-1MP | Dispersion | $2\alpha \sin(\frac{1}{2}\kappa) + (\frac{3}{2} - \frac{175}{128}\alpha)\sin(\kappa) +$ $(-\frac{3}{10} + \frac{7}{32}\alpha)\sin(2\kappa) + (\frac{1}{30} - \frac{3}{128}\alpha)\sin(3\kappa)$ |
| | Dissipation | $(\frac{2}{15} - \frac{3}{32}\alpha)\cos^3(\kappa) + (-\frac{2}{5} + \frac{9}{32}\alpha)\cos^2(\kappa) +$ $(\frac{2}{5} - \frac{9}{32}\alpha)\cos(\kappa) + (-\frac{2}{15} + \frac{3}{32}\alpha)$ |
| WENOIU5-2MP | Dispersion | $\frac{38}{15}\sin(\frac{1}{2}\kappa) - \frac{2}{15}\sin(\frac{3}{2}\kappa) - \frac{2}{15}\sin(\kappa) + \frac{1}{30}\sin(2\kappa)$ |
| | Dissipation | $-\frac{4}{15}\sin(\frac{1}{2}\kappa)^2(\cos(\frac{1}{2}\kappa) - 1)^2$ |
| WENOIU7-1MP | Dispersion | $2\alpha \sin(\frac{1}{2}\kappa) + \left(-\frac{735}{512}\alpha + \frac{8}{5}\right)\sin(\kappa) + \left(\frac{147}{512}\alpha - \frac{2}{5}\right)\sin(2\kappa) +$ $\left(-\frac{27}{512}\alpha + \frac{8}{105}\right)\sin(3\kappa) + \left(\frac{5}{1024}\alpha - \frac{1}{140}\right)\sin(4\kappa)$ |

|  | Dissipation | $(-\frac{2}{35} + \frac{5}{128}\alpha)\cos^4(\kappa) + (\frac{8}{35} - \frac{5}{32}\alpha)\cos^3(\kappa) +$ <br> $(-\frac{12}{35} + \frac{15}{64}\alpha)\cos^2(\kappa) + (\frac{8}{35} - \frac{5}{32}\alpha)\cos(\kappa) + (-\frac{2}{35} + \frac{5}{128}\alpha)$ |
|---|---|---|
| CS4 | Dispersion | $\frac{9}{4}\sin(\frac{1}{2}\kappa) - \frac{1}{12}\sin\left(\frac{3}{2}\kappa\right)$ |
| CS6 | Dispersion | $\frac{189}{80}\sin(\frac{1}{2}\kappa) - \frac{25}{192}\sin\left(\frac{3}{2}\kappa\right) + \frac{3}{320}\sin\left(\frac{5}{2}\kappa\right)$ |

4.2 ADR analysis

It is known that the spectral properties of nonlinear schemes will differ from their linear counterparts, and the properties will appear differently case by case. As a compromise, an Approximate Dispersion Relation method (ADR) was proposed by [40], where the spectral characteristics of schemes were numerically derived by means of solving advection problems. Currently, the linear advection problem in Section 5.1 is chosen for ADR analysis. The initial complex distribution at the domain [0, 1] with periodic boundary is: $u(x,t=0) = 0.01e^{i\kappa_n}$, where $\kappa_n = (2\pi n)/N$ with $n=0...N/2$ and $N=122$. The computation advances over a tiny period as $\tau=4.0\times10^{-6}$, where the third-order TVD-RK3 scheme [2] is chosen for temporal discretization and aforementioned PPM techniques are employed. The result of the dispersion and dissipation relations are shown in Fig. 2, which are classified into two groups similar to that in Section 4.1 except that the dispersion relation of CS6 is added in the first group.

From Fig. 2a, WENOIU3 schemes show dissipations larger than that of CS4, where WENOIU3-2MP behaves the most dissipative. Regarding the dispersion, WENOIU3-1MP shows the similar performance as that of CS4, while WENOIU3-2MP demonstrates an enhanced one comparable to that of CS6.

From Fig. 2b, the spectral distributions of schemes show a trend which is qualitatively similar to results with interpolations incorporated in Fig. 1b, i.e. all dispersion relations are similar to each other while WENOIU5-1MP shows the least dissipative.

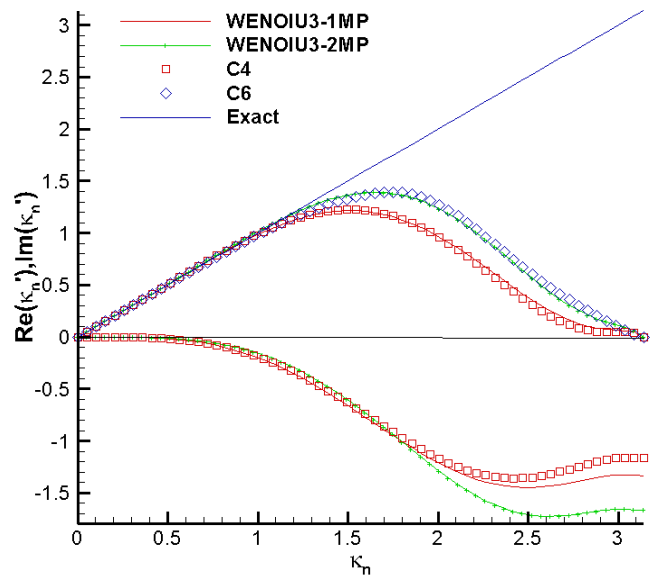

(a) WENOIU3-1MP, -2MP, CS4 and CS6

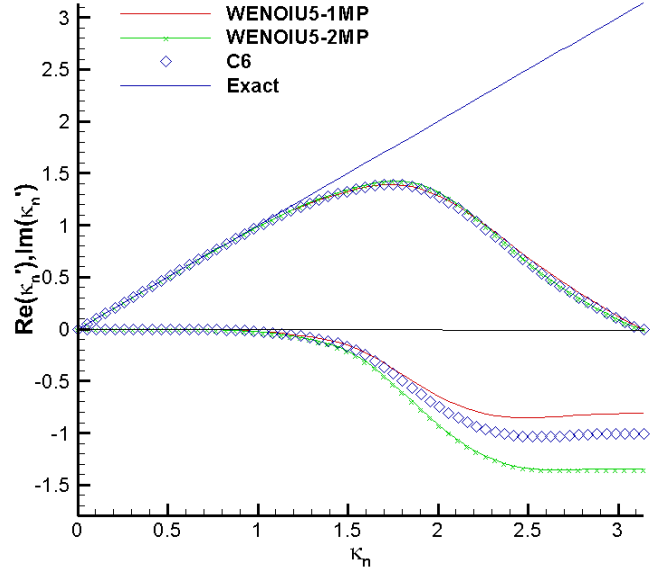

(b) WENOIU5-1MP, -2MP and CS6

Fig.2 Dispersion and dissipation relations of difference schemes by ADR

**5 Tests on convergence order and free-stream preservation**

5.1 Convergence order by computations of linear advection

The linear advection equation is

$$u_t + u_x = 0,$$

where $x \in [-1,1]$ and uniformed grids are used. Standard fourth-order Runge-Kutta scheme was used to discretize the temporal term, and the time step is chosen as $(\Delta x)^{5/4}$ to minimize possible influence of temporal errors. The case of smooth initial distribution is chosen for test as

$$u(x,t=0) = \sin(\pi x - \sin(\pi x)/\pi).$$

The computation is advanced to $t=2$. In Table 3, numerical errors and corresponding orders are shown regarding proposed schemes, namely, WENIOU3-1MP, WENIOU3-2MP, WENIOU5-1MP, and WENIOU5-2MP, where aforementioned PPM functions are employed. A special treatment is chosen for WENOIU3 schemes similar to that in [39], i.e. $IS_0$ and $IS_1$ in Eq. (3.22) on three-point stencils are employed instead of that in Eq. (3.14) in the same sequence. It is easy to verify that the new use of $IS$s satisfy the accuracy requirement of weighted schemes, which manifests their qualification as indicators in the weighting procedure. The employment of new $IS$s comes from the fact that Eq. (3.14) fails to recover the third-order order accuracy no matter what kinds of mapping functions are used (e.g. the one in WENO-M [5], the third-, fourth- or even the fifth-order PPM functions [7]), while the linear counterpart of interpolation shown in Table 1 can make the order achieved. The cause is suspected to be the inherent inability of two-point stencil to describe the distribution with extremum, and the occurrence is planned to discuss in other studies. It can be seen from Table 3 that the designed orders are well established.

**Table 3. Convergence properties of proposed schemes in linear advection problem with smooth initial condition**

| N | WENOIU3-1MP | | WENOIU3-2MP | | WENOIU5-1MP | | WENOIU5-2MP | |
|---|---|---|---|---|---|---|---|---|
| | $L_2$ (order) | $L_\infty$ (order) | $L_2$ (order) | $L_\infty$ (order) | $L_2$ (order) | $L_\infty$ (order) | $L_2$ (order) | $L_\infty$ (order) |
| 10 | 0.12 (----) | 0.20 (----) | 0.10 (----) | 0.19 (----) | 5.67E-02 (----) | 0.11 (----) | 5.31E-02 (----) | 0.11 (----) |
| 20 | 2.20E-02(2.46) | 4.47E-02(2.18) | 1.75E-02(2.56) | 3.71E-02(2.36) | 2.96E-03(4.26) | 6.24E-03(4.16) | 2.58E-03(4.36) | 5.43-E03(4.29) |
| 40 | 3.64E-03(2.59) | 7.78E-03(2.52) | 2.81E-03(2.64) | 6.09E-03(2.61) | 1.06E-04(4.81) | 2.10E-04(4.89) | 8.94E-05(4.84) | 1.73-E04(4.97) |
| 80 | 5.01E-04(2.86) | 1.05E-03(2.88) | 3.78E-04(2.89) | 7.87E-04(2.95) | 3.22E-06(5.03) | 6.70E-06(4.97) | 2.69E-06(5.05) | 5.48-E06(4.98) |
| 160 | 6.42E-05(2.96) | 1.28E-04(3.03) | 4.82E-05(2.97) | 9.61E-05(3.03) | 9.84E-08(5.03) | 2.10E-07(5.00) | 8.12E-08(5.05) | 1.71-E07(5.00) |
| 320 | 8.08E-06(2.99) | 1.60E-05(3.00) | 6.06E-06(2.99) | 1.20E-05(3.00) | 3.04E-09(5.01) | 6.55E-09(5.00) | 2.49E-09(5.02) | 5.35-E09(5.00) |
| 640 | 1.01E-06(3.00) | 2.00E-06(3.00) | 7.59E-07(3.00) | 1.50E-06(3.00) | 9.46E-11(5.01) | 2.05E-10(5.00) | 7.71E-11(5.02) | 1.67-E10(5.00) |

In the following, $IS_0$ and $IS_1$ by Eq. (3.22) are used for WENOIU3 schemes instead of Eq. (3.14) in order to achieve desired accuracy.

5.2 Free-stream preservation on randomized grid

An initial uniformed flow condition is employed with the Mach number as 0.5 on the randomized grids described in Section 6.1. Using the proposed third- and fifth-order schemes, the computation runs until $t=10$ with the time step $\Delta t=0.01$. $L_2$ errors of velocity component $v$ are shown in Table. 4. In the computation, flux splitting uses the aforementioned Lax-Friedrichs-type scheme. From the table, the free-stream preservation is well achieved by WENOIU schemes on randomized grid.

**Table 4. $L_2$ errors of $v$ component in *FSP* test on randomized grids**

| | WENOIU3-1MP | WENOIU3-2MP | WENOIU5-1MP | WENOIU5-2MP |
|---|---|---|---|---|
| $L_2(v)$ | 1.820E-15 | 1.500E-15 | 2.401E-15 | 2.512E-15 |

**6 Numerical examples**

In all examples, Euler equations are solved with the employment of PPM methods for interpolations in WENOIU and with the choice of third-order TVD-RK3 method [2] for temporal discretization. Firstly, the grid generations are introduced regarding subsequent 2-D computations.

6.1 Grid generations in 2-D problems

(1) Grids for vortex preservation

Two grids are employed, namely the uniformed and randomized grids. The latter is generated by:

$$x_{i,j} = -\frac{L}{2} + \frac{L}{I_{max}-1}\left[(i-1) + 2A_{i,j}\left(Rand(0,1) - 0.5\right)Rand(0|1)\right]$$

$$y_{i,j} = -\frac{L}{2} + \frac{L}{J_{max}-1}\left[(j-1) + 2A_{i,j}\left(Rand(0,1) - 0.5\right)(1 - Rand(0|1))\right]$$

where $L=16$, $A_{ij} = 0.45$ at $i=5...I_{max}-4$ or $j=5...J_{max}-4$ otherwise $A_{ij} = 0$, Rand(0, 1) is a random function ranging from 0 to 1 while Rand(0|1) is one having the value 0 or 1. In this study, $(I_{max} \times J_{max})=(81\times81)$. The grids are shown in Fig. 3 where the initial vorticity contours are displayed as the background. It is worth mentioning that the grid randomization is so serious that further increase of $A_{ij}$ will cause negative grid-cell area. Besides the use in vortex preservation, the randomized grids are also utilized in the test of free-stream preservation in Section 5.2.

The uniformed grids have the same dimensions and physical range as that of the randomized ones.

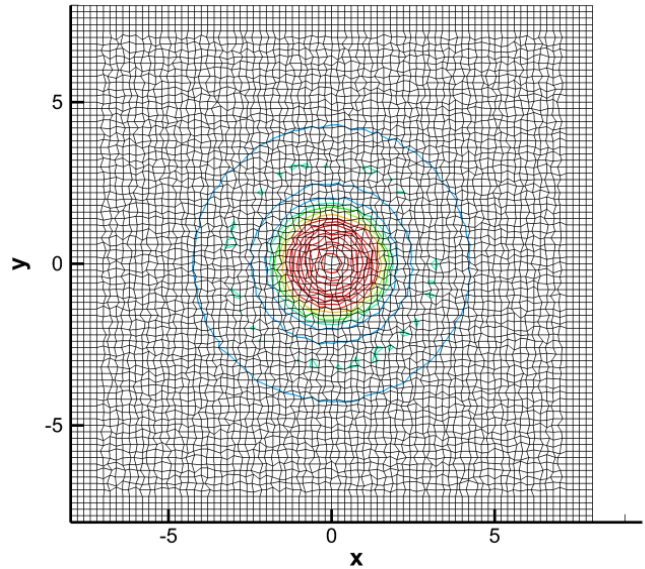

Fig. 3 81×81 randomized grids of vortex preservation with background by initial vorticity contours

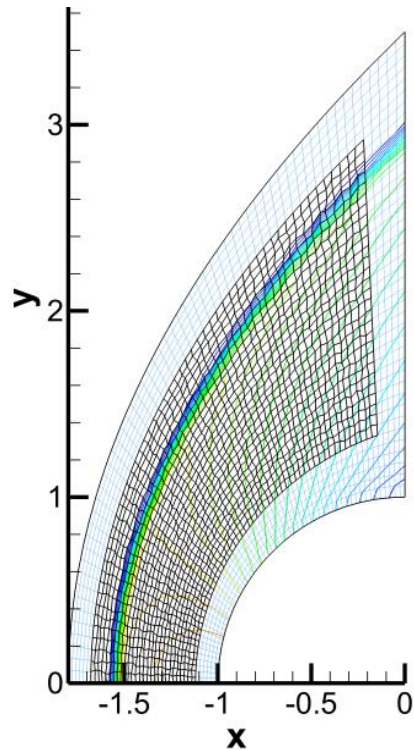

Fig. 4 121×41 grids of cylinder with background by pressure contours with randomization in a window

(2) Grids for supersonic flow around inviscid cylinder

A cylinder with the diameter being one is considered. Similarly, the smooth grids and the randomized ones are generated. The grid dimension ($I_{max} \times J_{max}$) is 121×41 which regards the streamwise and normal-wise directions. The smooth grids are generated by ordinary transfinite interpolation, while the randomized ones are acquired by randomizing the grids in a window of the former as that in "(1)". The indices of window is (6, $I_{max}$ - 5)× (7, $J_{max}$ - 6), and the randomizing factor $A_{ij}$ is 0.2. In Fig. 4, the half of referred window together with grids are shown with the background by pressure contours. From the figure, it can be seen that the shock wave and some flow after it are contained in region with randomized grids.

(3) Grids for Riemann problem and shock-vortex interaction

The physical domains are both rectangular in two cases. The range of domain in Riemann problem is [0, 1] × [0, 1] with 400 × 400 grids, and the one in shock-vortex interaction is [-20, 10] × [-15, 15] with 600 × 600 grids. Two types of grids are chosen, i.e. the uniformed grids and the randomized ones. For the latter, the grids are generated just as that in "(1)", where the factor $A_{ij}$ for inner randomized grids is 0.2, and where the four and nine layers of uniformed grids are engaged in boundaries respectively in Riemann problem and shock-vortex interaction.

6.2 One-dimensional problems

(1) Sod problem

The initial distributions of the problem at the range [-5, 5] is

$$(\rho, u, p) = \begin{cases} (0.125, \ 0, \ 0.1) & -5 \leq x < 0 \\ (1.000, \ 0, \ 1.0) & 0 \leq x \leq 5 \end{cases},$$

and the computation is advanced to t=2.0 on 100 uniformed grids with $\Delta t$=0.01. On checking, WENOIU3 schemes yield results quite similar to each other, and so do WENOIU5 schemes. Hence for clarity, only the density distribution of WENOIU3-1MP and WENOIU5-2MP are shown as representatives in Figs. 5 and 6. From the figures, the fifth-order scheme indicates relatively better performance at the corners of contact discontinuity than that of the third-order scheme.

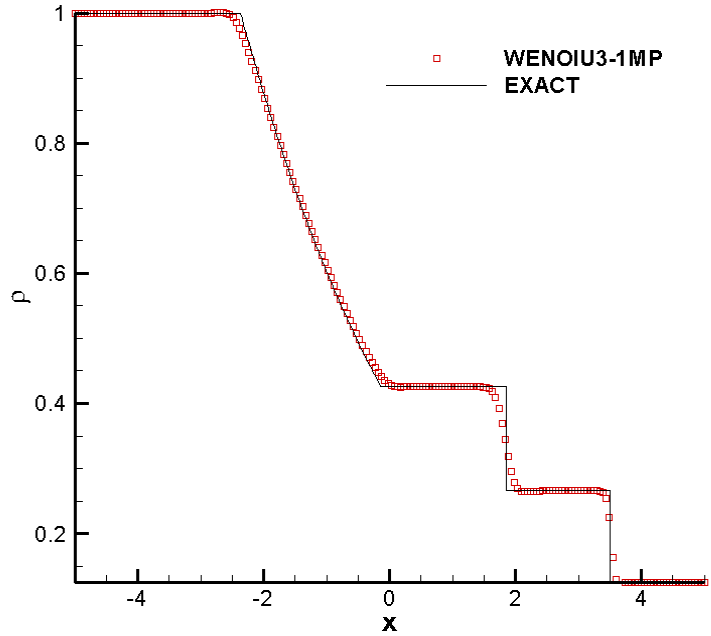

Fig.5 Density distribution of Sod problem by WENOIU3-1MP

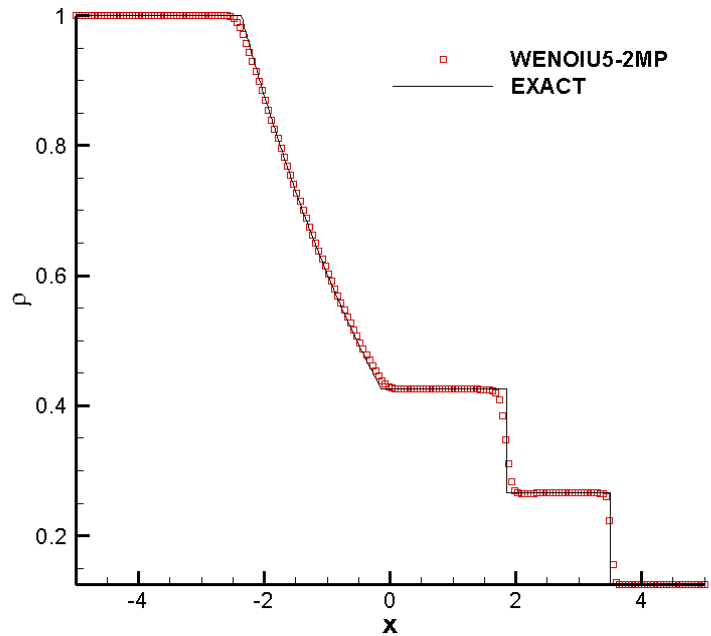

Fig.6 Density distribution of Sod problem by WENOIU5-2MP

(2) Shu-Osher problem

The initial distributions of the problem at the range [-5, 5] is

$$(\rho,u,p) = \begin{cases} (3.857143, 2.629369, 10.3333) & x < -4 \\ (1+0.2\sin(5x), 0, 1) & x \geq -4 \end{cases},$$

and the computation is advanced to t=1.8 on 400 uniformed grids with $\Delta t$=0.001. WENOIU3 schemes show similar results to each other again and only that of WENOIU3-2MP is shown in Fig. 7 for clarity. One can check that the result shows better resolution if compared with canonical WENO3 [2], and the improvement should be attributed to the new use of *IS* discussed before. In

Fig. 8, WENOIU5 schemes show well descriptions on the small-scale structures after the shock, where WENOIU5-2MP yields similar result to that of WENOIU5-1MP and therefore is omitted for brevity.

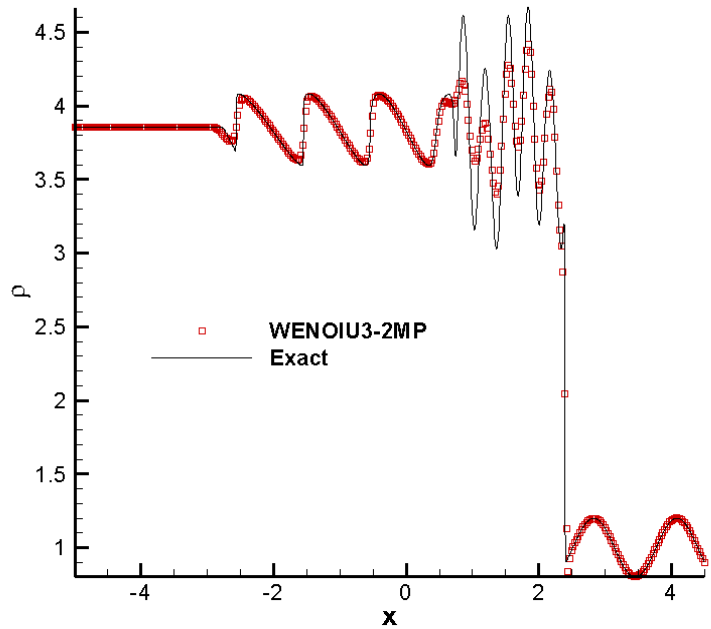

Fig.7 Density distribution of WENOIU3-2MP for Shu-Osher problem

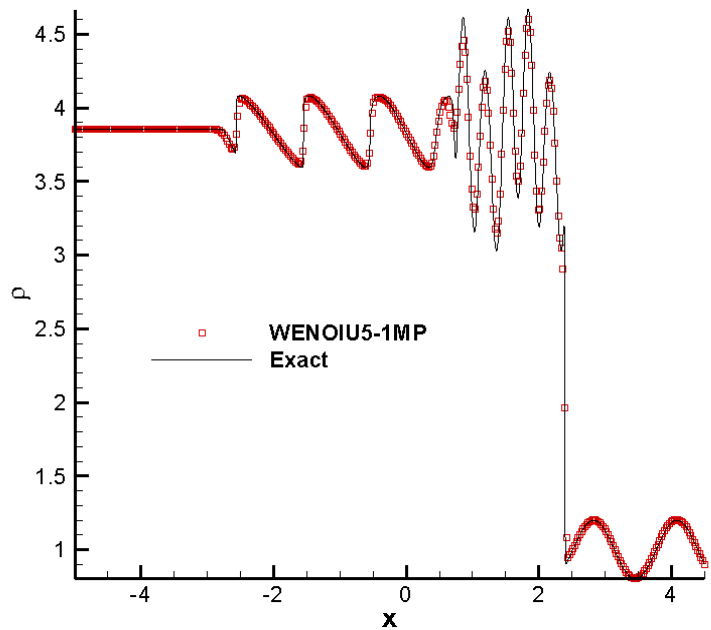

Fig.8 Density distribution of WENOIU5-1MP for Shu-Osher problem

6.3 Two-dimensional problems

(1) Vortex preservation

This problem is to check if a moving vortex could preserve its shape in grids introduced in Section 6.1 with periodic boundaries. The flow is non-dimensionalized by the density and the speed of sound, and the free-stream Mach number is 1. An isentropic vortex is initially superimposed on the uniformed flow at $\vec{r}_0 = (0, 0)$ as [24]

$$\begin{cases} (\delta u, \delta v) = \varepsilon \tilde{r} e^{\alpha(1-\tilde{r}^2)}(\sin\theta, -\cos\theta) \\ \delta T = -\dfrac{(\gamma-1)\varepsilon^2}{4\alpha\gamma} e^{2\alpha(1-\tilde{r}^2)} \\ \delta S = \delta(p/\rho^\gamma) = 0 \end{cases},$$

where $\tilde{r} = |\vec{r} - \vec{r}_0|/r_c$, $r_c=1$, $\alpha=0.204$, $\varepsilon=0.3$, and $\gamma=1.4$. The computation runs from the initial conditions at $\Delta t=0.01$ till $t=16$, which corresponds to one movement circle of the vortex to return to its initial place through the periodic boundary.

WENOIU3 and WENOIU5 schemes are respectively tested in uniformed and randomized grids. In the tests of WENOIU3 schemes, WENOIU3-2MP shows quite similar result to that of WENOIU3-1MP, therefore the vorticity contours of the latter are omitted here. Similarly, only the contours of WENOIU5-2MP are shown for representative.

On the uniformed grids shown in Fig. 9a, WENOIU3-1MP yields a result with quite smooth appearance and well circular shape, and the quantitative comparison of WENOIU3 with the exact solution is fairly well as shown in Fig. 10a. On randomized grids (Fig. 9b), WENOIU3-1MP yields a result with reasonable preservation of vortex, which is comparable to that in [24] and is consistent to the quantitative distributions in Fig. 10b. It is worthwhile to mention that the vorticity contours are especially chosen for visualization other than isobar contours, and the consideration is that the latter usually have smoother appearance which might conceal potential problems. Besides, it is reminded that on randomized grids, even the initial vorticity contours do not appear smoothly enough due to visualizations as shown in Fig. 3.

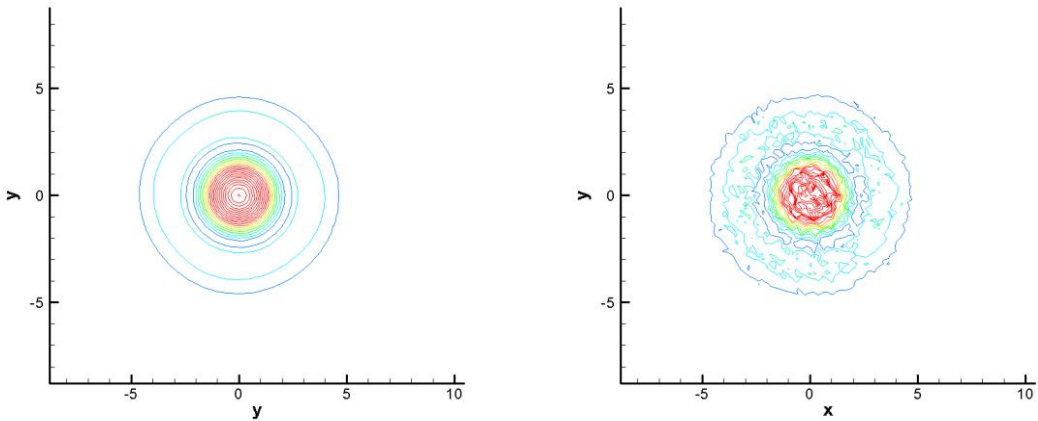

(a) Results on uniformed grids   (b) Results on randomized grids

Fig. 9 Vorticity contours of vortex preservation problem on two 81×81 grids by using WENOIU3-1MP (Contours from 0 to 0.7 with number 21)

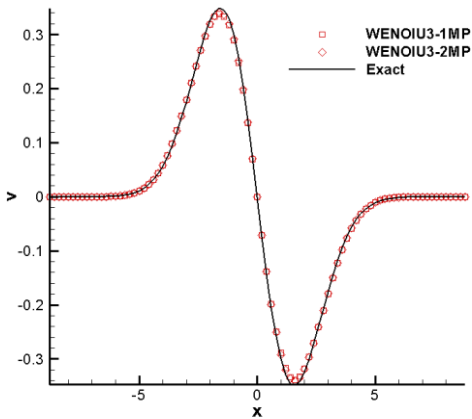 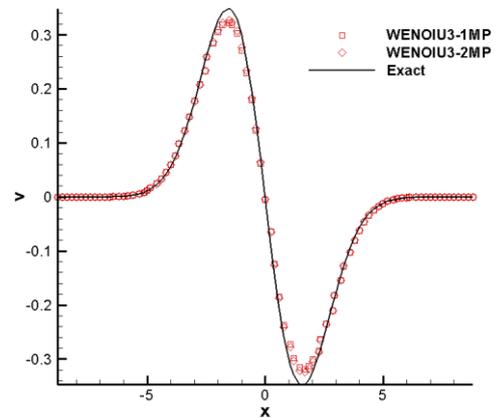

(a) Results on uniformed grids      (b) Results on randomized grids

Fig. 10 Distributions of *v*-component along the line at $j = \frac{Jmax}{2} + 1$ of vortex preservation problem on two 81×81 grids by using WENOIU3 schemes

The results of WENOIU5-2MP are shown in Figs. 11 and 12, and from the figures the structure of vortex is found to be qualitatively similar to that of WENOIU3-1MP. Quantitatively, WENOIU5-2MP behaves less dissipatively and shows slightly better agreement with the exact distribution.

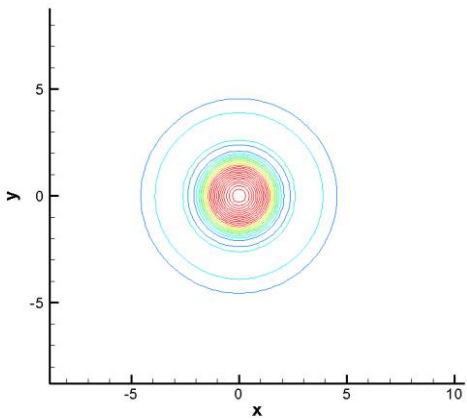 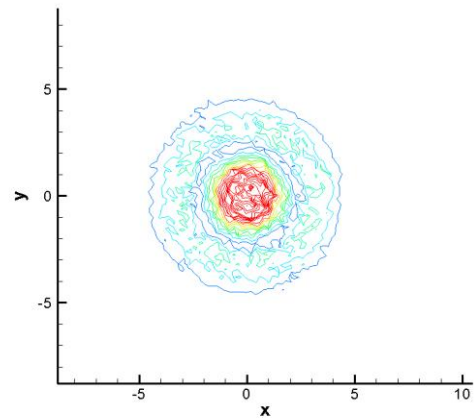

(a) Results on uniformed grids      (b) Results on randomized grids

Fig. 11 Vorticity contours of vortex preservation problem on two 81×81 grids by using WENOIU5-2MP (Contours from 0 to 0.7 with number 21)

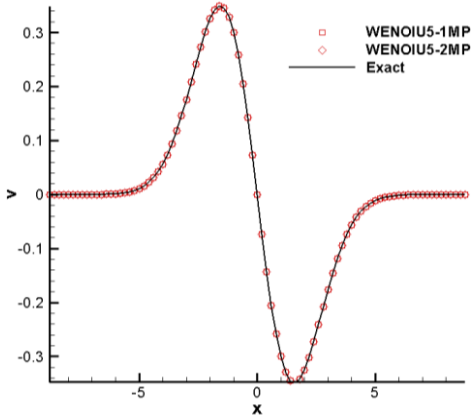 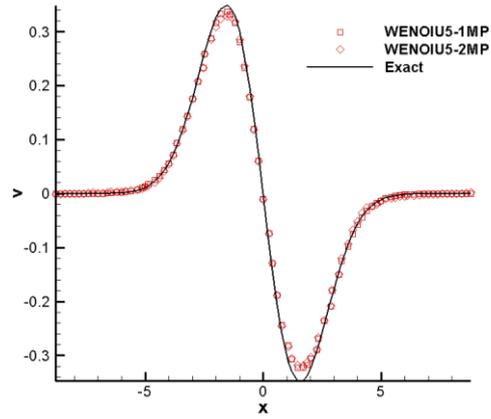

(a) Results on uniformed grids          (b) Results on randomized grids

Fig. 12 Distributions of *v*-component along the line at $j = \frac{Jmax}{2} + 1$ of vortex preservation problem on two 81×81 grids by using WENOIU5 schemes

(2) Supersonic inviscid flow around the cylinder at $M_\infty=4$

The computation is carried out at $\Delta t=0.001$ and advanced till $t=400$. First, the proposed third order WENOIU3 schemes are tested. On both grids, the schemes produce qualitatively similar results, therefore the results of two schemes on two grids are selectively shown in Fig. 13. In the meanwhile, quantitatively checking shows that the parametric distributions of WENOIU3-2MP occasionally have tiny overshoots after the shock wave. From the figure, the capability of proposed methods for resolving shock wave on grids with bad quality is manifested. More carefully checking shows that on randomized grids, the contours near the shock appear relatively unsmooth when compared with the ones on smooth grids, and similar situations occur in the flow after the shock. Such features are also observed in subsequent tests, which indicates the influence of grid quality still exists.

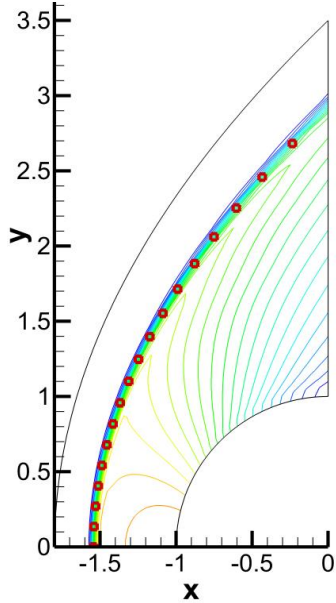 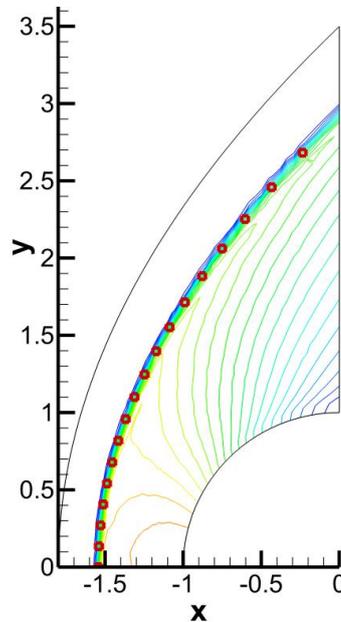

(a) Results of WENOIU3-1MP on uniformed grids      (b) Results of WENOIU3-2MP on randomized grids

Fig. 13 Pressure contours of supersonic cylinder flow on two 121×41 grids by using WENOIU3 schemes (Contours from 1.2 to 5.2 with number 21; Dots: solution from Lyubinov& Rusanov[41])

Next, the results of WENOIU5 schemes are investigated as well. Similar characteristics are observed as that in computations by using WENOIU3. Qualitatively, the contours on randomized grids appear relatively unsmooth than that in Fig. 13b, which might arise from the less dissipation in the fifth-order schemes.

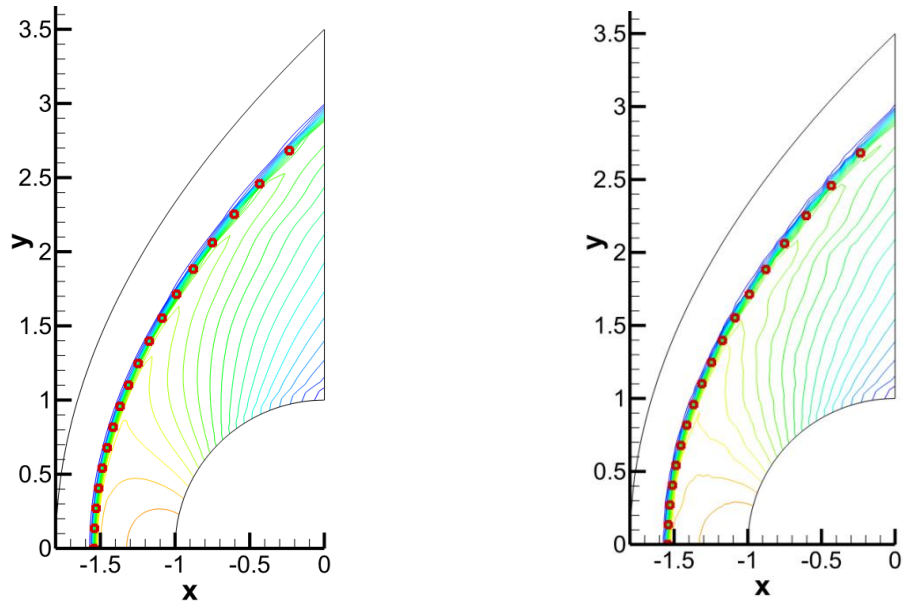

(a) Results of WENOIU5-2MP on uniformed grids

(b) Results of WENOIU5-1MP on randomized grids

Fig. 14 Pressure contours of supersonic cylinder flow on two 121×41 grids by using WENOIU5 schemes (Contours from 1.2 to 5.2 with number 21; Dots: solution from Lyubinov& Rusanov[41])

To quantitatively check the numerical predictions, the pressure distributions along wall surface of WENOIU3-2MP and WENOIU5-1MP on randomized grids are depicted in Fig. 15 with the comparison with the asymptotic solution by Lyubinov and Rusanov [41]. From the figure, well agreements are obtained between the computations and the asymptotic solution.

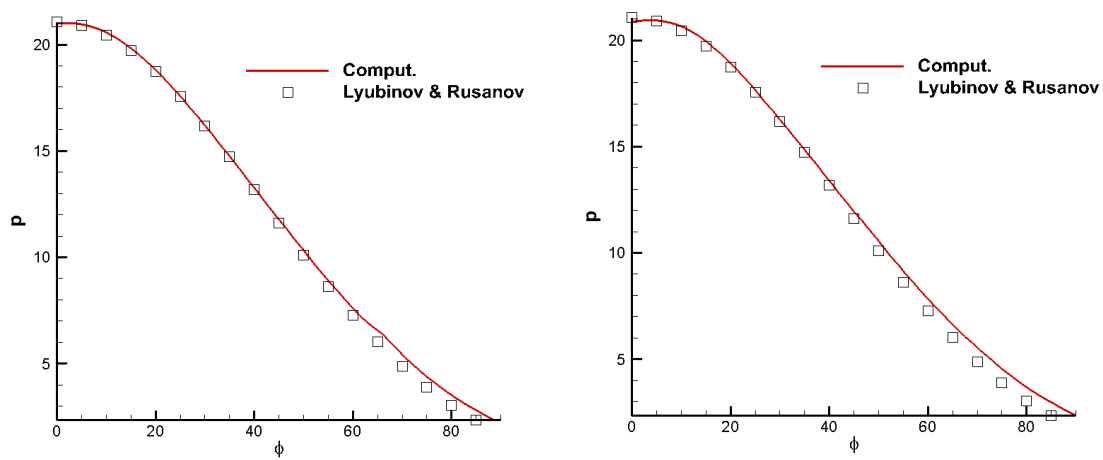

(a) Results of WENOIU3-2MP on uniformed grids

(b) Results of WENOIU5-1MP on randomized grids

Fig. 15 Pressure distributions along wall surface of supersonic cylinder flow on 121×41 randomized grids by WENOIU3-2MP and WENOIU5-1MP (Dots: solution from Lyubinov and Rusanov [41])

(3) Riemann problem

The following initial condition is imposed for four sub-divisions of the domain $[0, 1]^2$ as:

$[0.5, 1] \times [0.5, 1]$: $\rho_1=1$, $p_1=1$, $u_1=0$, $v_1=-0.3$;

$[0, 0.5] \times [0.5, 1]$: $\rho_2=2$, $p_2=1$, $u_2=0$, $v_2=0.3$;

$[0, 0.5] \times [0, 0.5]$: $\rho_3=1.0625$, $p_3=0.4$, $u_3=0$, $v_3=0.8145$;

$[0.5, 1] \times [0, 0.5]$: $\rho_4=0.5313$, $p_4=0.4$, $u_4=0$, $v_4=0.4276$.

The Euler equation is solved on 400×400 grids introduced previously, and the computations proceed to $t=0.3$. Two time steps are employed, i.e. $\Delta t=0.001$ for uniformed grids and $\Delta t=0.0001$ for randomized grids. The reason to use a smaller step in the latter is because the numerical stability is decreased there and only the reduction of time step can fulfill the computation. The results of WENOIU3 schemes are first discussed. On uniformed grids, the result of WENOIU3-1MP is quite similar to that of WENOIU3-2MP, therefore only the latter is shown in Fig. 16a for representative. On randomized grids, WENOIU3-1MP fulfills the computation with the result shown in Fig. 16b while WENOIU3-2MP fails to complete the job even if $\Delta t$ is largely decreased. Hence, WENOIU3-1MP is able to yield smooth density contours on both uniformed and randomized grids, which manifests the capability to resolve shock and slip line on grids with bad quality. In the meanwhile, in Fig. 16b the theoretically un-changed flow-fields preserve their profiles while continuously-varied fields show smooth distributions, which implies the achievement of *FSP* by WENOIU3-1MP. The failure of WENOIU3-2MP on randomized grids indicates its weaker numerical stability.

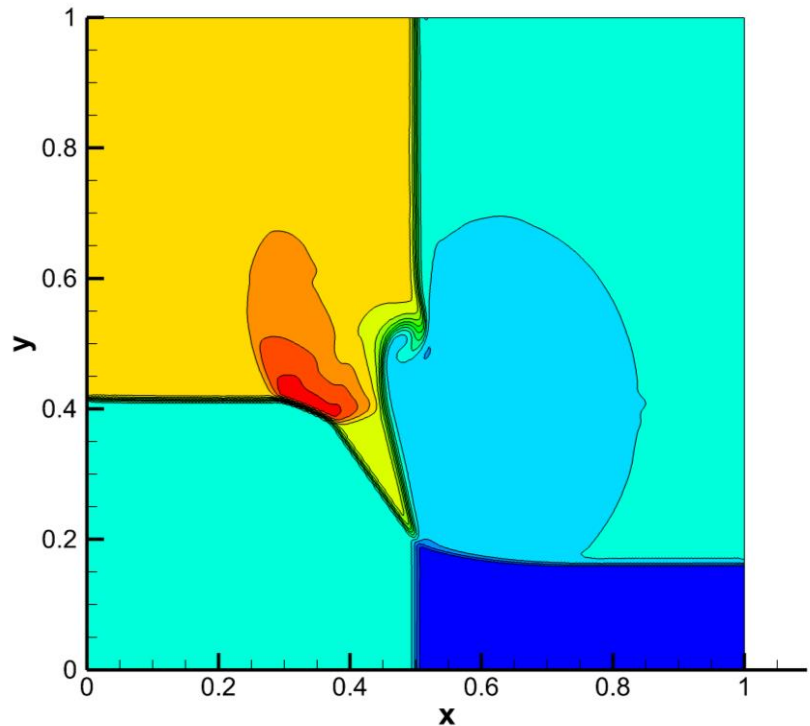

(a) Results of WENOIU3-2MP on uniformed grids

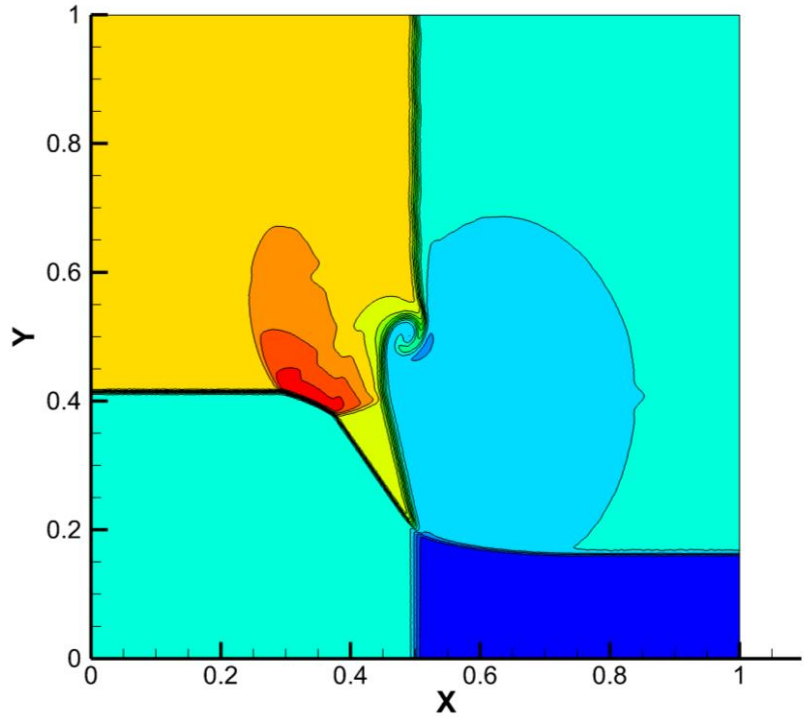

(b) Results of WENOIU3-1MP on randomized grids

Fig. 16 Density contours of Riemann problem on two 401×401 grids by using WENOIU3 schemes (Contours from 0.6 to 2.3 with number 14)

Next, the results of WENOIU5 schemes are checked and shown in Fig. 17. On uniformed grids, WENOIU5-1MP yields quite smooth result while WENOIU5-2MP produces a result with some oscillations. Although some subtle structures are generated at the slip line by the latter, their generation is suspected of the lack of enough numerical stability. On randomized grids, both schemes fulfill the computations and yield similar results. If compared with that by WENOIU3 schemes, WENOIU5 schemes indicate better resolutions on description of the slip line.

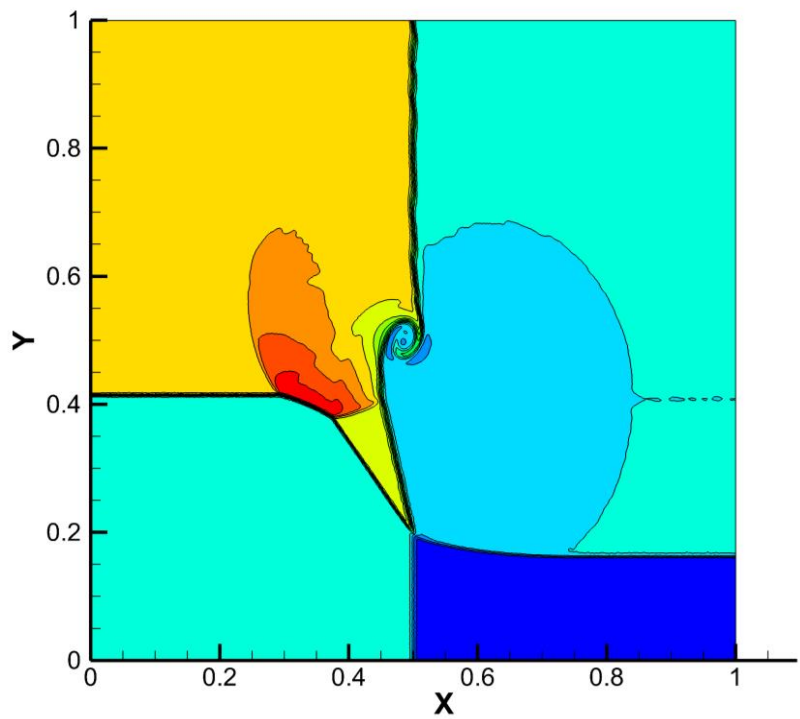

(a) Results of WENOIU5-1MP on uniformed grids

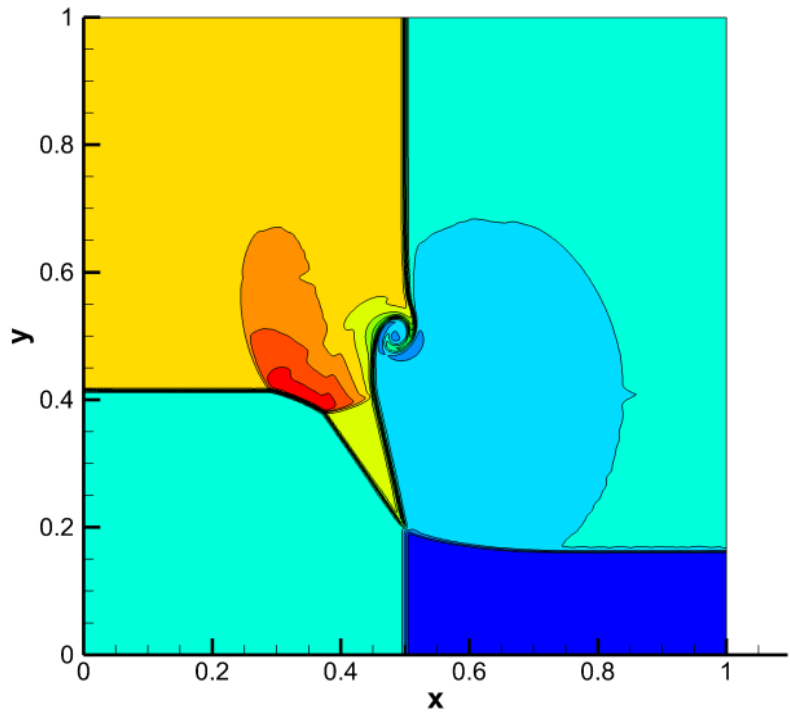

(b) Results of WENOIU5-1MP on randomized grids

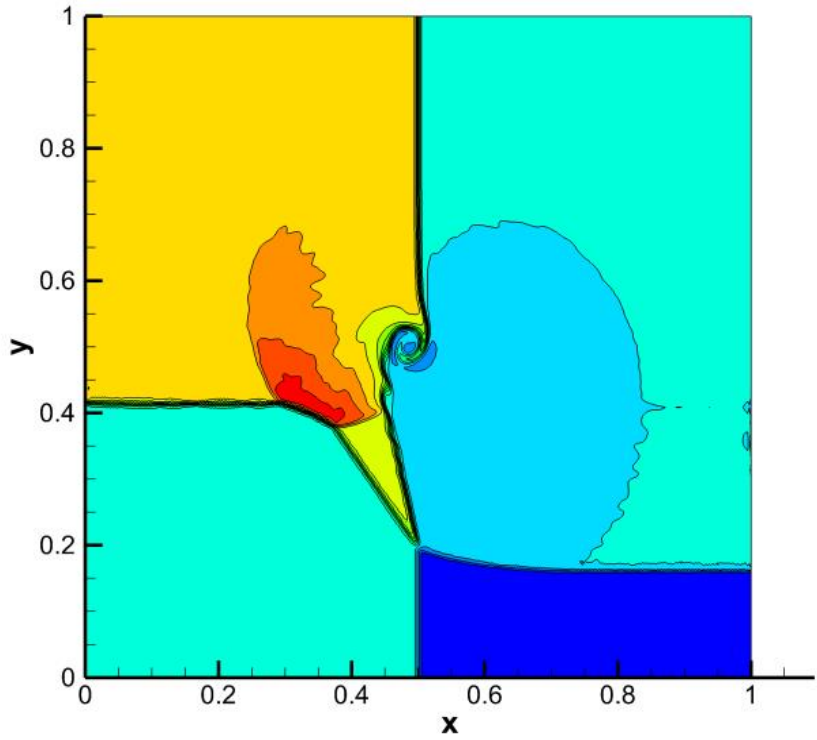

(c) Results of WENOIU5-2MP on uniformed grids

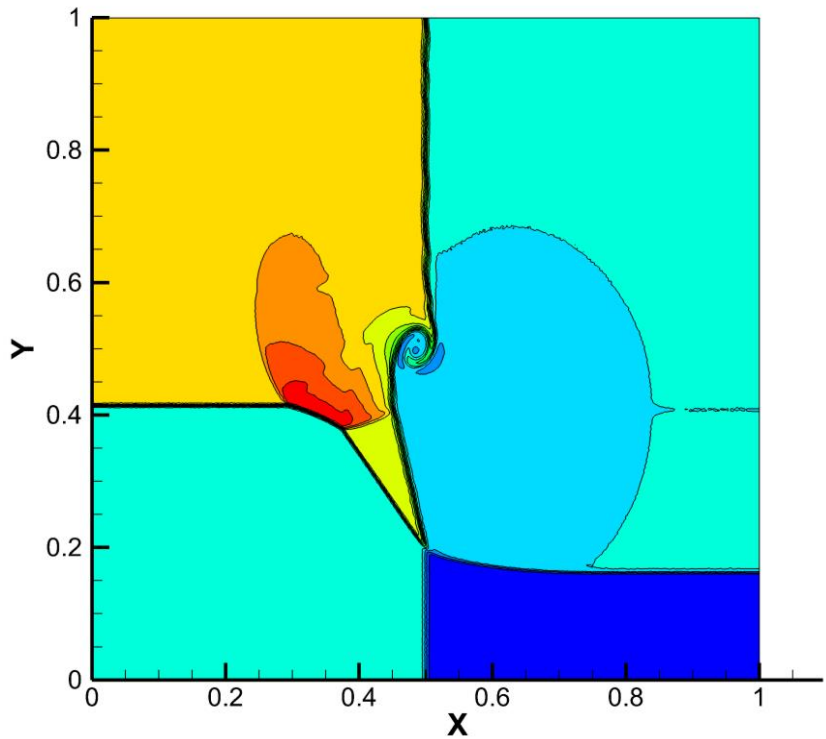

(d) Results of WENOIU5-2MP on randomized grids

Fig. 17 Density contours of Riemann problem on two 401×401 grids by using WENOIU5 schemes (Contours from 0.6 to 2.3 with number 14)

(4) Shock-vortex interaction

The case regards a strong shock with $M_s=3$ interacting with a strong vortex. The computational domain is: [-20, 10]×[-15, 15], where the shock wave is initially set and fixed at $x=0$ by boundary conditions. The initial flow-field after the shock can be obtained by Rankine - Hugoniot relations. An isentropic vortex is superimposed ahead of the shock wave as:

$$u_\theta = \Gamma |\vec{r} - \vec{r}_0|/r_c \, e^{\left[(1-|\vec{r}-\vec{r}_0|^2)/2\right]},$$

$$p(r) = \frac{1}{\gamma M_\infty^2}\left[1 - \frac{\gamma-1}{2}\Gamma^2 \exp(1-|\vec{r}-\vec{r}_0|^2)\right]^{\gamma/(\gamma-1)},$$

$$\rho(r) = \left[1 - \frac{\gamma-1}{2}\Gamma^2 \exp(1-|\vec{r}-\vec{r}_0|^2)\right]^{1/(\gamma-1)},$$

where $u_\theta$ denotes the tangential velocity of the vortex, $\vec{r}_0 = (4,0)$ represents the initial position of the vortex, and $\Gamma$ is the vortex strength with the value 0.4. The computation runs to $t=15$ with $\Delta t=0.01$.

The results of WENOIU3 schemes are checked first. On uniformed grids, both WENOIU3-1MP and -2MP schemes yield results with no qualitative difference, therefore only the density contours of WENOIU3-2MP are drawn in Fig. 18a for representative. On randomized grids, WENOIU3-1MP works normally with the result shown in Fig. 18b, while WENOIU3-2MP blows up even if $\Delta t$ is largely decreased. This consequence seems to indicate again that WENOIU3-2MP be less stable when the grids are of poor quality.

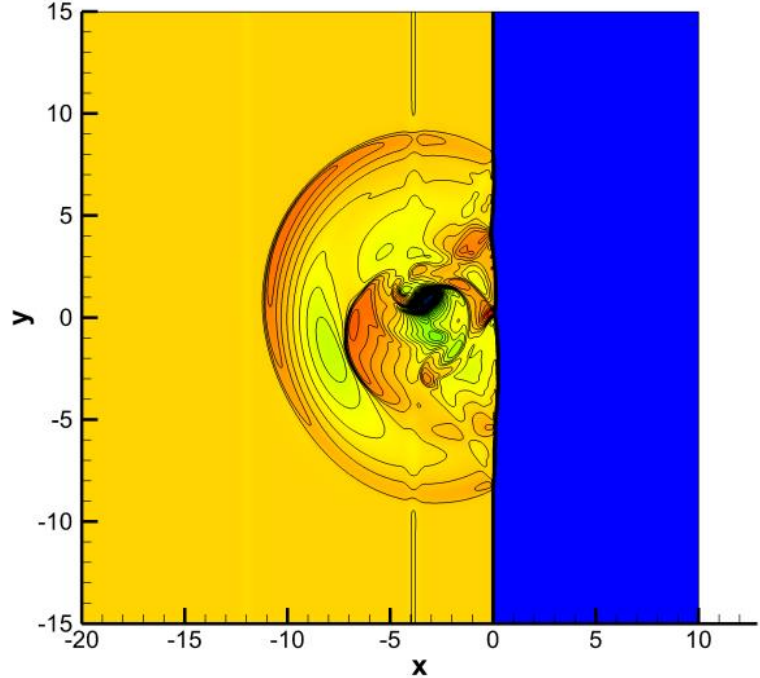

(a) Results of WENOIU3-2MP on uniformed grids

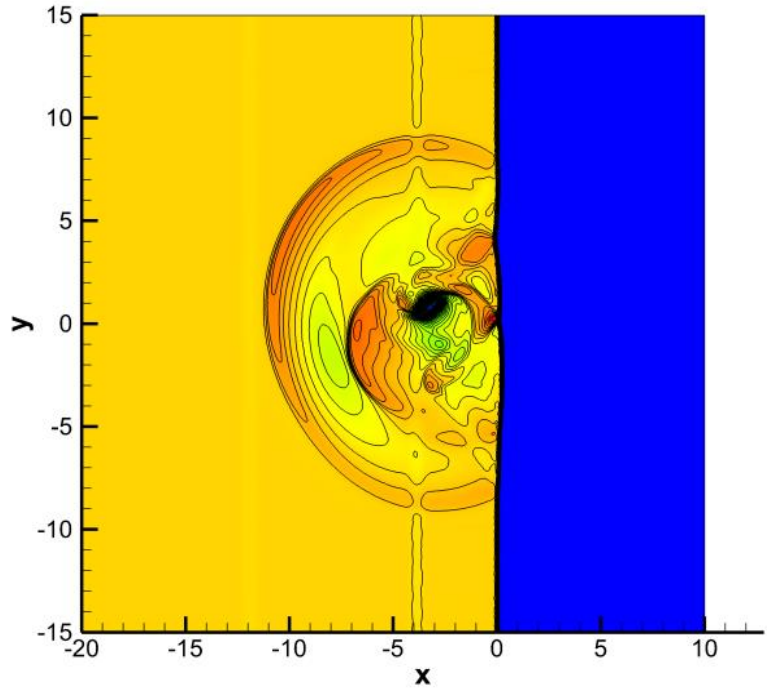

(b) Results of WENOIU3-1MP on randomized grids

Fig. 18 Density contours of shock-vortex interaction on two 601×601 grids by using WENOIU3 schemes (Contours from 0.6 to 2.3 with number 14)

The results of WENOIU5 schemes are shown in Fig. 19. Because results of WENOIU5-1MP and -2MP schemes show no difference qualitatively on uniformed grids, only the density contours of WENOIU5-2MP are shown. On randomized grids, both schemes fulfill the computations, and the result of WENOIU5-1MP is shown in Fig. 19b as an example. From Figs 19a and 19b, although the structures by randomized grids resemble to that on uniformed grids, the former appears relatively oscillatory than the latter, e.g. small oscillations are observed after the shock. In addition, although WENOIU5-2MP yields a result with similar structures as that of WENOIU5-1MP, carefully checking shows that the contours by WENOIU5-2MP appear more oscillatory. Hence, WENOIU5-2MP might have relatively inferior performance when the grids are of bad quality.

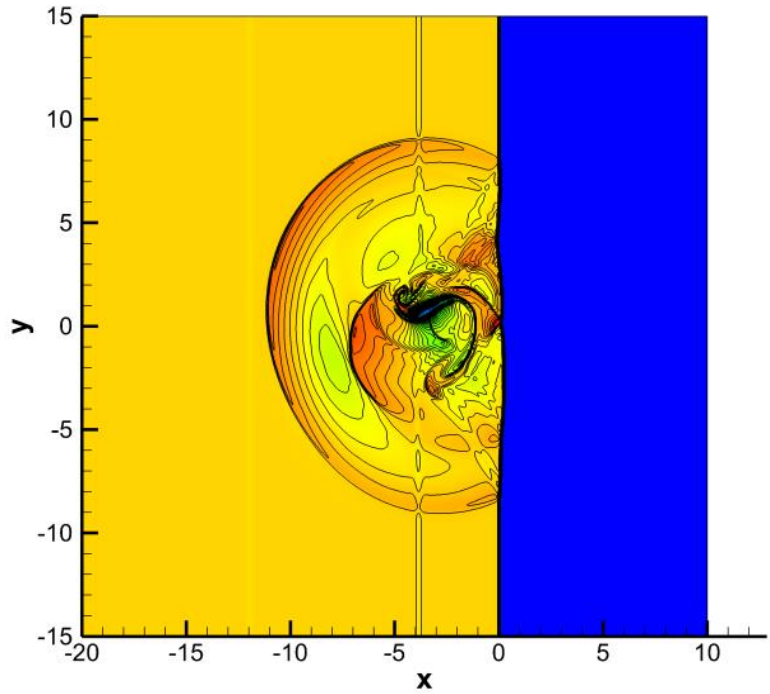

(a) Results of WENOIU5-2MP on uniformed grids

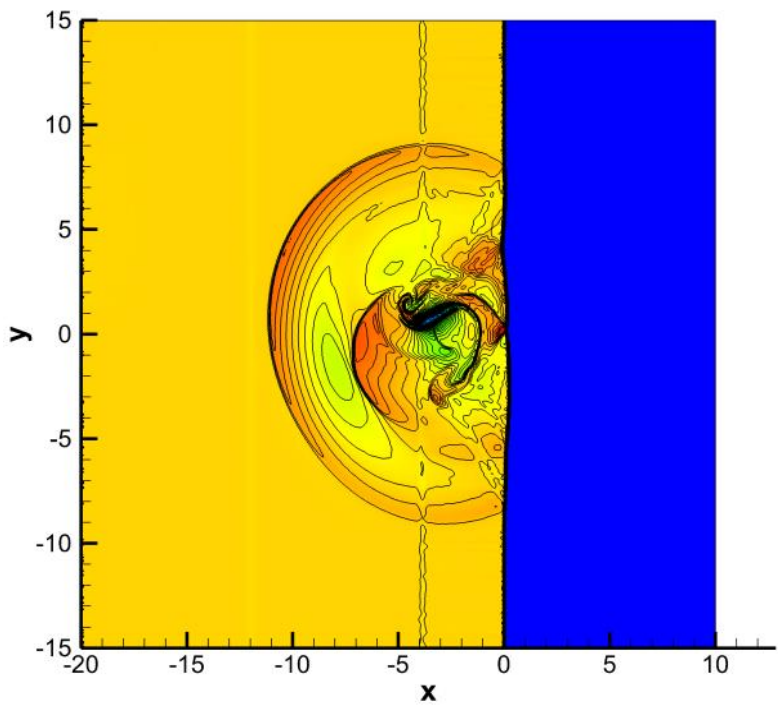

(b) Results of WENOIU5-1MP on randomized grids

Fig. 19 Density contours of shock-vortex interaction on two 601×601 grids by using WENOIU5 schemes (Contours from 0.6 to 2.3 with number 14)

**7 Concluding remarks and discussions**

Current studies are the nonlinear extensions of the work in [26] where linear upwind schemes with flux splitting were proposed with the achievement of *FSP*. By means of engagement of midpoints and WENO interpolations, nonlinear upwind-biased schemes are developed with *FSP* achieved, while the third- and fifth-order ones are tested by typical examples. The following conclusions are drawn:

(1) Tests of linear convection on different grids show that all WENOIU3 and WENOIU5 schemes can achieve their designed order. To obtain the objective, *PPM* should be casted in nonlinear interpolations, and in addition, the smoothness indicators in WENOIU3 should choose the ones on three-point stencils other than those on two-point stencils.

(2) Tests of *FSP* on randomized grids show that all WENOIU3 and WENOIU5 schemes can achieve *FSP* property, therefore the methodology discussed in the paper succeeds in making nonlinear upwind-biased schemes to realize this important property. The fulfillment of vortex preservation tests further validate the above statement.

(3) Three problems with shock waves are chosen to test WENOIU3 and WENOIU5 schemes, where randomized/partially randomized grids are used and complex flow structures exist sometimes. All schemes fulfill the case of supersonic cylinder flow, and nearly all of them complete the Riemann problem and shock-vortex interaction except WENOIU3-2MP fails on randomized grids. The results on the one hand manifest the capability of proposed schemes for solving problems on grids with bad quality, on the other hand the relative weak stability of WENOIU3-2MP is also indicated. Generally speaking, WENOIU schemes with two midpoints seems to be less robust or more oscillatory than ones with one midpoint.

One can see from Eqns. (3.4), (3.8), (3.17) and (3.20) that $f_{j+1/2}$ takes a special position in proposed schemes. Actually for the computation of $f_{j+1/2}$, different flux splitting schemes can be employed, which will not influence the achievement of *FSP*. Hence the frequently-used schemes with low dissipation can be employed, e.g. the Steger-Warming scheme, Van Leer scheme, and etc. In the meanwhile, the recent advances in WENO-Z schemes can be applied in interpolations as well. Such issues should deserve attentions in further studies.

**Acknowledgements.**

This work is sponsored by the National Science Foundation of China under the Grant Number 91541105, and the second author also thanks for the support of the National Science Foundation of China under the Grant Number 11802324. The corresponding author thanks for the contribution of Dr. Qilong Guo on the incipient 1-D computations.

**Appendix**

For completeness, formulations of the seventh-order WENOIU schemes are provided in a similar way as that in Section 3.2 and 3.3. As before, only upwind-biased schemes are concerned.

(1) Scheme with only one midpoint $f_{j+1/2}$ engaged

The general form with one free parameter is

$$h_{j+1/2} = \alpha f_{j+1/2} + \left(\tfrac{5}{1024}\alpha - \tfrac{1}{140}\right)f_{j-3} + \left(-\tfrac{21}{512}\alpha + \tfrac{5}{84}\right)f_{j-2} + \left(\tfrac{175}{1024}\alpha - \tfrac{101}{420}\right)f_{j-1} + \\ \left(-\tfrac{175}{256}\alpha + \tfrac{319}{420}\right)f_{j} + \left(-\tfrac{525}{1024}\alpha + \tfrac{107}{210}\right)f_{j+1} + \left(\tfrac{35}{512}\alpha - \tfrac{19}{210}\right)f_{j+2} + \left(-\tfrac{7}{1024}\alpha + \tfrac{1}{105}\right)f_{j+3}.$$

When $\alpha=1$, the scheme becomes

$$h_{j+1/2} = f_{j+1/2} + \tfrac{1}{107520}\begin{pmatrix} -243f_{j-3} + 1990f_{j-2} - 7481f_{j-1} + 8164f_{j} \\ -341f_{j+1} - 2378f_{j+2} + 289f_{j+3} \end{pmatrix}.$$

The corresponding eighth-order $\delta^{c,(1)}f$ is

$$\delta^{c,(1)}f = \frac{1}{\Delta x}\left(f_{j+1/2} - f_{j-1/2}\right) + \frac{1}{107520\Delta x}\begin{bmatrix}-\frac{243}{2}\left(f_{j+4}-f_{j-4}\right)+1261\left(f_{j+3}-f_{j-3}\right)\\-6069\left(f_{j+2}-f_{j-2}\right)+8841\left(f_{j+1}-f_{j-1}\right)\end{bmatrix}.$$

(2) Scheme with two midpoints engaged

$$h_{j+1/2} = f_{j+1/2} + \frac{1}{1260}\begin{bmatrix}\left(-128f_{j-1/2}+404f_{j+1/2}\right)+\\\left(-2f_{j-2}+47f_{j-1}-92f_{j}-233f_{j+1}+5f_{j+2}\right)\end{bmatrix}.$$

The corresponding eighth-order $\delta^{c,(1)}f$ is

$$\delta^{c,(1)}f = \frac{1}{315\Delta x}\left[-16\left(f_{j+3/2}-f_{j-3/2}\right)+432\left(f_{j+1/2}-f_{j-1/2}\right)\right] +$$
$$\frac{1}{1260\Delta x}\left[-\left(f_{j+3}-f_{j-3}\right)+27\left(f_{j+2}-f_{j-2}\right)-189\left(f_{j+1}-f_{j-1}\right)\right].$$

(3) Scheme with four midpoints engaged

$$h_{j+1/2} = f_{j+1/2} + \frac{1}{420}\begin{bmatrix}\left(18f_{j-1}-66f_{j}-108f_{j+1}\right)+\\\left(-3f_{j-3/2}-31f_{j-1/2}+179f_{j+1/2}+11f_{j+3/2}\right)\end{bmatrix}.$$

The corresponding $\delta^{c,(1)}f$ is

$$\delta^{c,(1)}f = \frac{1}{140\Delta x}\left[3\left(f_{j+2}-f_{i-2}\right)-32\left(f_{j+1}-f_{j-1}\right)\right] +$$
$$\frac{1}{840\Delta x}\left[-3\left(f_{j+5/2}-f_{i-5/2}\right)-17\left(f_{j+3/2}-f_{j-3/2}\right)+1218\left(f_{j+1/2}-f_{j-1/2}\right)\right].$$

(4) Scheme with six midpoints engaged

$$h_{j+1/2} = f_{j+1/2} + \frac{1}{26880}\left[3840f_{j-1} + \begin{pmatrix}-45f_{j-5/2}+501f_{j-3/2}-3874f_{j-1/2}\\+746f_{j+1/2}-1249f_{j+3/2}+81f_{j+5/2}\end{pmatrix}\right].$$

The corresponding $\delta^{c,(1)}f$ is

$$\delta^{c,(1)}f = \frac{1}{14\Delta x}\left(f_{j+1}-f_{j-1}\right) + \frac{1}{\Delta x}\begin{bmatrix}-\frac{3}{3584}\left(f_{j+7/2}-f_{j-7/2}\right)+\frac{209}{17920}\left(f_{j+5/2}-f_{j-5/2}\right)\\-\frac{163}{1536}\left(f_{j+3/2}-f_{j-3/2}\right)+\frac{575}{512}\left(f_{j+1/2}-f_{j-1/2}\right)\end{bmatrix}.$$

WENO interpolations by Eqns. (3.10)-(3.11) are integrated to evaluate variables on midpoints, where the coefficients in the equations are shown in Table A1 [32].

**Table A1 Coefficients $b_{k,l}^r$ and $C_k^r$ in interpolation schemes**

| r | k | $C_k^r$ | $b_{k,l}^r$ | | | |
|---|---|---|---|---|---|---|
| | | | l=0 | l=1 | l=2 | l=3 |
| 4 | 0 | 1/64 | -5/16 | 21/16 | -35/16 | 35/16 |
| | 1 | 21/64 | 1/16 | -5/16 | 15/16 | 5/16 |
| | 2 | 35/64 | -1/16 | 9/16 | 9/16 | -1/16 |
| | 3 | 7/64 | 5/16 | 15/16 | -5/16 | 1/16 |

When order degradation is concerned, the following fourth-order PPM can be employed [7] to dispose the nonlinear weights in interpolation

$$g(\omega) = \begin{cases} C_k \left[1 - \left(\frac{\omega}{C_k} - 1\right)^4\right] & \omega \leq C_k \\ C_k - \left(\frac{1}{C_k - 1}\right)^3 (\omega - C_k)^4 & \omega > C_k \end{cases}.$$